\documentclass[a4paper,fleqn,twocolumn]{aastex63}

\usepackage{float}
\usepackage{subfiles}
\usepackage{graphicx}
\usepackage{enumitem}
\usepackage{amssymb}
\usepackage{amsmath}
\usepackage{hyperref}
\usepackage{aas_macros}
\usepackage[lofdepth,lotdepth,caption=false]{subfig}
\def\muhz{\mu\mathrm{Hz}}
\def\numax{\nu_{\mathrm{max}}}

\def\dnu{\Delta \nu}

\def\teff{T_{\mathrm{eff}}}

\def\teffsun{T_{\mathrm{eff,} \odot}}

\def\rsun{R_{\odot}}
\def\deg{^{\circ}}

\def\numaxsun{\nu_{\mathrm{max,} \odot}}
\def\dnusun{\Delta\nu_{\odot}}

\newcommand{\mas}{\rm{mas}}
\newcommand{\muas}{\mu\rm{as}}

\def\varpigaia{\varpi_{Gaia}}
\def\hatvarpigaia{\hat{\varpi}_{Gaia}}

\def\muhz{\mu\mathrm{Hz}}
\def\numax{\nu_{\mathrm{max}}}

\def\dnu{\Delta \nu}

\def\teff{T_{\mathrm{eff}}}

\def\teffsun{T_{\mathrm{eff,} \odot}}

\def\rsun{R_{\odot}}
\def\deg{^{\circ}}

\def\numaxsun{\nu_{\mathrm{max,} \odot}}
\def\dnusun{\Delta\nu_{\odot}}

\def\varpiast{\varpi_{\mathrm{seis}}}

\def\muas{\mu \mathrm{as}}
\def\mas{\mathrm{mas}}
\def\nueff{\nu_{\mathrm{eff}}}

\def\ks{K_{\mathrm{s}}}

\def\nueffunit{\mu \mathrm{m} ^{-1}}
\def\dunit{\muas\,\mu\mathrm{m}}
\def\daunit{\mu\mathrm{m}}
\def\zf{Z_5}

\graphicspath{{plots/}{../plots/}}
\begin{document}
\submitjournal{ApJ}
\shorttitle{Gaia EDR3 zero-point}
\shortauthors{J.~C. Zinn}
\title{Validation of the Gaia Early Data Release 3 parallax zero-point
  model with asteroseismology}
\correspondingauthor{Joel C. Zinn}
\email{jzinn@amnh.org}

\author{Joel C. Zinn}
\altaffiliation{NSF Astronomy and Astrophysics Postdoctoral Fellow.}
\affiliation{Department of Astrophysics, American Museum of Natural History, Central Park West at 79th Street, New York, NY 10024, USA}

\begin{abstract}
The Gaia Early Data Release 3 (EDR3) provides trigonometric parallaxes for 1.5 billion stars, with reduced systematics compared to Gaia Data Release 2 and reported precisions better by up to a factor of two. New to EDR3 is a tentative model for correcting the parallaxes of magnitude-, position-, and color-dependent systematics for five- and six-parameter astrometric solutions, $Z_5$ and $Z_6$. Using a sample of over 2,000 first-ascent red giant branch stars with asteroseismic parallaxes, I perform an independent check of the $Z_5$ model in a Gaia magnitude range of $9 \lesssim G \lesssim 13$ and color range of $1.4\nueffunit \lesssim \nueff \lesssim 1.5\nueffunit$. This analysis therefore bridges the Gaia team's consistency check of $Z_5$ for $G > 13$, and indications from independent analysis using Cepheids of a $\approx 15\muas$ over-correction for $G < 11$. I find an over-correction sets in at $G \lesssim 10.8$, such that $Z_5$-corrected EDR3 parallaxes are larger than asteroseismic parallaxes by $15 \pm 3 \muas$. For $G \gtrsim 10.8$, EDR3 and asteroseismic parallaxes in the Kepler field agree up to a constant consistent with expected spatial variations in EDR3 parallaxes after a linear, color-dependent adjustment. I also infer an average under-estimation of EDR3 parallax uncertainties in the sample of $22 \pm 6\%$, consistent with the Gaia team's estimates at similar magnitudes and independent analysis using wide binaries. Finally, I extend the Gaia team's parallax spatial covariance model to brighter magnitudes ($G < 13$) and smaller scales (down to $\approx 0.1\deg$), where systematic EDR3 parallax uncertainties are at least $\approx 3-4\muas$.
\end{abstract}

\section{Introduction}
\label{sec:intro}
The Gaia mission has provided astrometric information for over 1.5 billion stars as part of Gaia Early Data Release 3 (EDR3), and which is largely complete down to $G \sim 21$ in uncrowded regions \citep{gaia-collaboration+2020}. This release successfully builds upon the previous Data Release 2 \citep[DR2; ][]{gaia+2016,gaia-collaboration+2018a}, with improvements to the mission's angular resolution, completeness, and astrometric precision. In particular, the reported parallax precision in EDR3 has improved by a factor of two from $\approx 40\muas$ for sources with $G < 15$ to $\approx 20\muas$.

Because of the nominal increase in the parallax precision, it is all the more important to understand systematic uncertainties in the Gaia parallaxes. Following indications of systematic errors in the Gaia Data Release 1 parallaxes from the Gaia team and from subsequent independent investigations \citep{michalik_lindegren&hobbs2015,gaia+2016,lindegren+2016,huber+2017,zinn+2017plxtgas,jao+2016,deridder+2016,davies+2017,stassun&torres2016b}, several studies investigated the Gaia zero-point in Gaia Data Release 2. Many of the studies investigated the level of the global offset, which is due to degeneracies between variations in the angular separation of the spacecraft's fields of view and the parallax zero-point of the astrometric solution \citep{butkevich+2017,lindegren+2018}. These studies required independent parallax estimates to compare the Gaia parallaxes against, and ranged from independent trigonometric parallaxes \citep{leggett+2018}; classical Cepheid photometric parallaxes \citep{riess+2018a,groenewegen+2018a}; RR Lyrae photometric parallaxes \citep{muraveva+2018a,layden+2019,marconi+2021}; open cluster/OB association isochronal parallaxes \citep{yalyalieva+2018,melnik_dambis2020,sun_q+2020}; very long baseline interferometric parallaxes \citep{kounkel+2018,bobylev2019}; statistical parallaxes \citep{schoenrich+2019a,muhie+2021}; and red giant asteroseismic parallaxes \citep{khan+2019,hall+2019a}. 

Apart from the $10-100\muas$ global offset inferred by the aforementioned studies, the Gaia team also identified position-, color- and magnitude-dependent trends in the Gaia zero-point, which were thought to result from Gaia's scanning pattern and CCD response \citep{lindegren+2018,arenou+2018}.  Independent studies subsequently confirmed similar trends \citep{zinn+2019zp,zinn+2019rad,leung_bovy2019,chan_bovy2020,fardal+2021}.
 
The global offset, position-, color-, and magnitude-dependent parallax systematics quantified in DR2 are also present, though to a lesser extent, in EDR3 \citep{l20a,l20b}. The reduction in systematics is a result of EDR3 benefitting from a new astrometric solution using 34 months of data as opposed DR2's 22 months; improvements in EDR3 to the Velocity error and effective Basic Angle Calibration (VBAC) model, which models the basic angle variations that contribute to the global offset component of the parallax zero-point; as well as improvements in the photometric image parameter determination; \citealt{rowell+2020}) and its iterative inclusion in the astrometric solution \citep{l20a}.

Whereas the Gaia team did not recommended a specific parallax zero-point correction in DR2, a model for the parallax zero-point has been provided for EDR3. This model is fit according to an iterative solution based ultimately on a sample of quasars distributed across the sky \citep{l20b}, for which the EDR3 parallaxes should be effectively zero. The resulting model, denoted $\zf$ or $Z_6$, depending on whether the parallax is from a five-parameter or six-parameter astrometric solution\footnote{The Gaia astrometry falls into three classes: stars with position information only (two-parameter solutions); stars with color information from DR2 of quality enough to correct for chromaticity effects (five-parameter solutions); and stars where the color is an additional free parameter in the astrometric solution (six-parameter solutions). The latter two solutions have distinct parallax zero-point properties \citep{l20a,l20b}.}, once subtracted from the raw EDR3 parallaxes, is meant to remove most magnitude-, color-, and position-dependent errors.

The Gaia team has checked the $\zf$ model using stars in the Large Magellanic Cloud (LMC), open cluster members, and wide binaries \citep{l20b,fabricius+2020}, finding good performance across a range of magnitude and color parameter space.  

Since the EDR3 release, independent analyses have begun to quantify what adjustments may be required of the $\zf$ model. \cite{riess+2020}, using a sample of classical Cepheids ($G \lesssim 11$, $1.35\nueffunit \lesssim \nueff \lesssim 1.6\nueffunit$), infer an offset of $-14 \pm 6\muas$ such that corrected Gaia parallaxes are larger than Cepheid photometric parallaxes. \cite{bhardwaj+2020}, appealing to blue RR Lyrae ($\nueff > 1.5\nueffunit$), find an offset with corrected Gaia parallaxes of $-25 \pm 5\muas$ in the same direction. \cite{stassun_torres2021}, using 76 bright, blue eclipsing binaries ($5 \lesssim G \lesssim 12$, $\nueff > 1.5\nueffunit$) show no statistically significant parallax residuals after correction according to the Gaia parallax model ($+15 \pm 18\muas$). Moreover, an analysis using photometric parallaxes of red clump stars has recently indicated parallax residuals for $G < 10.8$ and as a function of ecliptic longitude \citep{huang+2021}.

There is, however, nearly a complete absence of stars with $1.4 \nueffunit< \nueff < 1.5\nueffunit, G < 13$ in either the LMC sample used for validation of the $\zf$ solution \citep{l20b} or any of the independent validation test samples thus far. This range in magnitude space is especially interesting to explore given that it links the $G > 13$ checks at similar colors with the LMC by the Gaia team to the $G< 11$ study from \cite{riess+2020}. I therefore consider in this paper the evidence for adjustments to the $\zf$ model by appealing to asteroseismic parallaxes of more than 2,000 first-ascent red giant branch stars, which occupy this region of magnitude and color parameter space. Specifically, I look for errors in Gaia parallaxes corrected according to the Gaia $\zf$ parallax zero-point model as a function of color and magnitude. In so doing, I also consider evidence for corrections to the formal statistical uncertainties in Gaia parallaxes, as well as evidence for spatially-correlated uncertainties in Gaia parallaxes.

\section{Data}
\label{sec:data}
Asteroseismic data are adopted from APOKASC-2 \citep{pinsonneault+2018}, which consists of asteroseismic parameters for red giant branch stars, $\numax$ and $\dnu$, derived from light curves acquired by the Kepler mission \citep{borucki+2010}, as well as spectroscopic temperatures and metallicities from the Apache Point Observatory Galactic Evolution
Experiment \citep[APOGEE;][]{majewski+2010}, a survey within Sloan Digital Sky Survey IV \citep{blanton+2017}. I also appeal to an independent analysis of Kepler data from \citep{yu+2018} using the SYD asteroseismic pipeline \citep{huber+2009}. Whereas the APOKASC-2 asteroseismic measurements are average measurements based on results from five asteroseismic pipelines, using data from SYD alone provides a check on the sensitivity of the result on the asteroseismic data.

I make use of APOGEE DR16 \citep{ahumada+2020} parameters in this work, updated from the APOGEE DR14 \citep{holtzman+2018} parameters provided in \citep{pinsonneault+2018}. Both APOGEE DR14 and DR16 data are taken using the $R = 22,500$ APOGEE spectrograph on the 2.5-m telescope of the Sloan Digital Sky Survey \citep{wilson+2019,gunn+2006}. The data reduction pipeline for APOGEE is described in \cite{nidever+2015}, and the spectroscopic analysis is performed using the APOGEE Stellar Parameter and Chemical Abundance Pipeline \citep[ASPCAP;][]{garcia_perez+2016}. The APOGEE temperatures are calibrated to be on the infrared flux method scale of \cite{ghb09} \citep{holtzman+2015}, which, for the red giant sample I work with here, implies temperatures are on an absolute scale to within $\approx 20K$ \citep{zinn+2019rad}; I adopt statistical uncertainties of $30K$.

I cross-matched the APOKASC-2 sample with EDR3 by first matching the APOKASC-2 stars to Gaia DR2 sources using the DR2-2MASS cross-match \citep{marrese+2019}, and then using the Gaia DR2 designations to match to EDR3 sources using the EDR3-DR2 cross-match \citep{torra+2020}. In so doing, I only kept stars that have angular separations between DR2 and EDR3 sources of less than $100\mas$.

Most of the APOKASC-2 stars have five-parameter solutions instead of six-parameter solutions (44 versus 2130 first-ascent red giant branch stars after quality cuts), so the following analysis only uses stars with five-parameter astrometry solutions, and thus concentrates of validation of the $\zf$ model.

To ensure that the astrometric solutions are not affected by binarity, I followed the quality cuts of \cite{fabricius+2020}, rejecting stars with \texttt{ruwe} $\geq 1.4$, \texttt{ipd\_frac\_multi\_peak} $ > 2$, and \texttt{ipd\_gof\_harmonic\_amplitude} $ \geq 0.1$.  \texttt{ipd\_frac\_multi\_peak} corresponds to the fraction of observations for which the source was identified as having two, resolved peaks in the image, and therefore is indicative of resolved binaries; \texttt{ipd\_gof\_harmonic\_amplitude} indicates the level of variation in the image goodness-of-fit as a function of scan direction, which, if large, would imply the image is asymmetric and therefore suggestive of an unresolved binary; \texttt{ruwe} indicates that the astrometric solution does not completely describe the motion of the source, and so can therefore identify unresolved binaries with variable photocenters \citep{l20a}.

I further restricted the sample to those with $G < 13$, which are observed within window classes WC0a and WC0b \citep{l20a}. These magnitude-defined windows define what pixel mask the source is read out with, and WC0 was divided into two for EDR3: sources with $11 \lesssim G \lesssim 13$ are observed with the WC0b window, and those with $G \lesssim 11$ with the WC0a window. Magnitude-dependent systematics in the astrometry may therefore be related to different behavior of photometry in the different window classes (see, e.g., Fig.~A.4 in \cite{l20a}). The $G < 13$ regime is further complicated, however, by the use of differing integration times through the use of `gates', and which themselves depend on magnitude. I will investigate the possibility of magnitude-dependent errors in corrected Gaia parallaxes in what follows.

I adopt \texttt{nu\_eff\_used\_in\_astrometry} (denoted $\nueff$ here) as a proxy for color, following \cite{l20b}, and which is defined based on DR2 photometry for the 5-parameter astrometry I use in this analysis \citep{l20a}; bluer stars have larger $\nueff$, and redder stars have smaller $\nueff$. I did not make cuts in color, since the sample lies well within the regime $ 1.24\nueffunit < \nueff < 1.72\nueffunit$; outside of this range, the image chromaticity point-spread function and line-spread function corrections are not calibrated, and so there is reason to believe they would suffer from a stronger color-dependent zero-point error \citep{l20b}. I will also fit for color-dependent terms to describe residual differences between asteroseismic and Gaia parallaxes in what follows.

The parallax of a star can be derived from the asteroseismic radius via the Stefan-Boltzmann law, written in the following form:
\begin{align}
  \label{eq:r_plx}
  \varpiast(\teff, F, R)&= F^{1/2} \sigma_{\mathrm{SB}}^{-1/2} \teff^{-2}
  R^{-1} \nonumber \\
  &= f_0^{1/2}10^{-1/5(m + BC_b(\teff) - A_b)} \sigma_{\mathrm{SB}}^{-1/2} \teff^{-2} R^{-1},
\end{align}
where $R$ is the asteroseismic radius (see below), $\teff$ is the effective temperature, $\sigma_{\mathrm{SB}}$ is the Stefan-Boltzmann constant, and $F$ is the stellar bolometric flux. Here, I have rewritten the flux in terms of a bolometric correction in photometric band, $b$, $BC_b$; the extinction that the passband, $A_b$; and a magnitude-flux conversion factor that assumes the solar irradiance of $f_0 = 1.361\times10^{6} \mathrm{erg\,s^{-1}\,cm^{-2}}$ \citep{mamajek+2015a}, and an apparent solar bolometric magnitude of $m_{\mathrm{bol}} = -26.82$ \citep{torres+2010a}. I work here with 2MASS $\ks$ photometry \citep{skrutskie+2006} to reduce extinction effects. I use a bolometric correction based on MIST models \citep{paxton+2011a,paxton+2013a,paxton+2015a,paxton+2018a,paxton+2019} as implemented in \texttt{isoclassify} \citep{huber+2017,berger+2020}. Visual extinctions were adopted from \cite{rodrigues+2014a}, and converted into $A_{\ks}$ assuming $A_{\ks} = 0.113 A_V$ \citep{schlafly_finkbeiner2011}. 

The asteroseismic radii required to yield parallaxes according to Equation~\ref{eq:r_plx}, can be computed according to a scaling relation,
\begin{equation}
\label{eq:scalingapo}
\frac{R}{\rsun} \approx \left(\frac{\numax}{\numaxsun}\right) \left(\frac{\dnu}{f_{\dnu}\dnusun}\right)^{-2}\ \left(\frac{\teff}{\teffsun}\right)^{1/2},
\end{equation}
where $\numax$ is an asteroseismic observable that corresponds to the frequency at which stochastically-excited oscillations in the stellar envelopes have their highest amplitude and $\dnu$ is an asteroseismic observable that describes the approximately constant separation in frequency between two oscillation modes of adjacent radial order but the same spherical degree. The former is theoretically and empirically related to stellar surface gravity \citep{brown+1991,kjeldsen&bedding1995,chaplin+2008,belkacem+2011}. The latter is related to the mean stellar density \citep{ulrich1986,kjeldsen&bedding1995}, though there is evidence that the observed $\dnu$ requires a correction \citep[e.g.,][]{white+2011,guggenberger+2016,sharma+2016}, which is represented in the above by $f_{\dnu}$. $\dnusun$ and $\numaxsun$ represent measurements of these quantities for the Sun, to which the scaling relations are tied. The solar reference values of $\dnusun = 135.146 \muhz$ and $\numaxsun = 3076 \muhz$ that I adopt from \cite{pinsonneault+2018} are calibrated such that asteroseismic masses agree with dynamical masses from eclipsing binaries in NGC 6791 \citep{grundahl+2008,brogaard+2011,brogaard+2012a} and NGC 6819 \citep{brewer+2016_ngc6819,jeffries+2013,sandquist+2013}. $f_{\dnu}$, also adopted from \cite{pinsonneault+2018}, amount to $\approx 3\%$ corrections of $\dnu$, and which vary on a star-by-star basis according to mass, surface gravity, temperature, and metallicity.

In this work, I also consider effectively correcting both $\dnu$ and $\numax$ using non-linear scaling relations from \cite{kallinger+2018}, which are derived from fits to dynamical surface gravities and mean stellar densities in addition to open clusters NGC 6791 and NGC 6819:
\begin{align}
\label{eq:scaling3}
     \frac{R}{\rsun} &\approx \left(\frac{\numax}{\nu_{\mathrm{max,ref}}}\right)^{\kappa} \left(\frac{\dnu}{\dnu_{\mathrm{ref}}}\right)^{-2} \nonumber \\
     &\left[1.0 - \gamma \left(\log_{10}\frac{\dnu}{\dnu_{\mathrm{ref}}}\right)^2\right]^{2}\ \left(\frac{\teff}{\teffsun}\right)^{1/2},
\end{align}
where $\kappa = 1.0075 \pm 0.0021$, $\gamma = 0.0043 \pm \pm 0.0025$, and I use the $\dnu_{\mathrm{ref}}$ value from \cite{kallinger+2018} appropriate to the method of deriving APOKASC-2 $\dnu$ values, $\dnu_{\mathrm{ref}} = 133.1 \pm 1.3 \muhz$. I adopt $\nu_{\mathrm{max,ref}} = 3076 \muhz$ \citep{pinsonneault+2018}. In what follows, I marginalize over errors in the asteroseismic radii due to, e.g., the choice of $\nu_{\mathrm{max,ref}}$  or $\dnu$ corrections, by fitting for a multiplicative factor that brings the radii into agreement with the Gaia parallaxes.

I conservatively restricted the analysis to stars with $R < 30\rsun$, based on findings from \cite{zinn+2019rad} that asteroseismic radii are likely inflated in the evolved red giant regime. I also restricted the analysis to the first-ascent red giant branch stars with $15\muhz < \numax < 200\muhz$ and $2\muhz < \dnu < 10 \muhz$, which is the range occupied by the majority of the open cluster calibrators used by \cite{pinsonneault+2018} and \cite{kallinger+2018} to set $\numax$/$\dnu$ solar reference values/corrections. For parts of the following analysis using the \cite{yu+2018} dataset, I use the non-linear scaling relations of Equation~\ref{eq:scaling3}, since the APOKASC-2 $\dnu$ corrections ($f_{\dnu}$ in Eq.~\ref{eq:scalingapo}) assume the $\numax$ and $\dnu$ of the APOKASC-2 catalogue. 

\section{Methods}
The approach I take to investigate the need, if any, of adjustments to the $\zf$ five-parameter solution parallax zero-point model, is to compare Gaia parallaxes corrected using $\zf$ to asteroseismic parallaxes. By taking the difference between the corrected Gaia parallaxes and the asteroseismic parallaxes, which I refer to as the parallax residuals in what follows, one can simultaneously constrain asteroseismic parallax problems due to asteroseismic radius and any additive problems in the Gaia parallax that may remain after correction according to $\zf$. The simultaneous calibration is possible because errors in the asteroseismic parallax due to the asteroseismic radius will be fractional and therefore dependent on parallax, given the multiplicative dependence of asteroseismic parallax on the asteroseismic radius (Eq.~\ref{eq:r_plx}). By contrast, Gaia parallax zero-point problems are found to be additive, and can be described by terms that depend on magnitude, position, and color \citep{l20b}.

Formally speaking, the model for describing any adjustments to the $\zf$ model can therefore be described as a likelihood of the form
\begin{align}
  \label{eq:second_model}
  \begin{split}
&\mathcal{L}(c,d,e,a,k,g,s | \hatvarpigaia, \teff,
    \dnu, \numax, {A}_{\ks},\\
    & \ks, {BC}, \nueff, {G}, \sin \beta) \propto  \\
&\frac{1}{\sqrt{(2 \pi)^N |C^{'}|}} \exp{\left[-\frac{1}{2} (\vec{y} -
  \vec{x})^{\mathrm{T}}  C^{'-1} (\vec{y} - \vec{x})\right]}.
\end{split}
\end{align}
In the above,
\begin{align*}
\vec{y} &\equiv Y(a,d_a,\nueff)\varpiast(\teff, \dnu, \numax, {A}_{\ks}, \ks, {BC}) 
      \end{align*}
      and
      \begin{align*}
\vec{x} &\equiv \hatvarpigaia  - \zf(\nueff, G, \sin\beta) +
\Delta Z(c,d,e,\nueff, G),
\end{align*}
where 
\begin{equation}
\label{eq:ast_corr}
    Y(a,d_a,\nueff) \equiv 1-a + d_a(\nueff -1.48)
\end{equation}
describes fractional errors in asteroseismic parallaxes: $a$ describes errors in the asteroseismic radius scale due to, e.g., solar reference value choice, and would be unity in the absence of any. As I discuss in \S\ref{sec:color}, the term $d_a$ describes color-dependent corrections to the asteroseismic parallaxes due to, e.g., bolometric correction systematics that are a function of temperature/color.

$\Delta Z$ is the model for the residual parallax difference left unexplained by the $\zf$ model. $\hatvarpigaia$ is the raw EDR3 parallax, $\sin\beta$ is the sine of ecliptic latitude, and $\zf$ is evaluated using the Python implementation of the correction, \texttt{zero\_point}\footnote{\url{https://gitlab.com/icc-ub/public/gaiadr3\_zeropoint/-/tree/master}}. I will only be using Gaia parallaxes corrected in this way in the analysis, which I denote $\varpigaia$, as opposed to the raw Gaia parallaxes, denoted $\hatvarpigaia$. 

I consider several different forms for $\Delta Z$, allowing for a
local offset to the corrected Gaia parallaxes as well as color- and
magnitude-dependent terms. Since I only analyze stars in the Kepler
field, it is not possible to constrain ecliptic latitude--dependent terms as the
Gaia team has for the $\zf$ model (though it is possible to statistically infer the level of small-scale variations as a function of angular scale, as I discuss below and in \S\ref{sec:spatial}). At its most complicated, the model for the parallax residuals takes the form
\begin{align}
\label{eq:dz}
\Delta Z(c,d,e,\nueff, G) = \nonumber\\ \begin{cases}
    &  c + c_2 + d_1(\nueff - 1.48)  \\
    & + d_3(1.48 - \nueff)^3 + e_1({G} - 12.2) - \Delta Z_{G=13},\\
    & \mathrm{for\ } G < 10.8\\
      & c + d_2(\nueff - 1.48) \\
       &+ d_3(1.48 - \nueff)^3 + e_2({G} - 12.2) - \Delta Z_{G=13},\\
      & \mathrm{for\ } G \geq 10.8. \\
    \end{cases}
\end{align}

I consider nested models under various permutations of the parameters, $c$, $d$, $e$, etc., as listed in Table~\ref{tab:res}; a blank entry indicates an unused parameter, such that it can be considered to be zero in Equation~\ref{eq:dz}. In some models, there is a single color term across all magnitudes, i.e., $d_1 \equiv d_2 \equiv d$. Models otherwise differ by removing any number of the terms by setting them to zero (e.g., the preferred model, Model 0, has a single color term and no magnitude term, i.e., $d_1 \equiv d_2 \equiv d$ and $e_1 \equiv e_2 \equiv e = 0$).  The term $\Delta Z_{G=13}$ is not a fitted parameter, but rather a constant defined such that $c$ describes the mean offset at $G=13$ and $\nueff = 1.48\nueffunit$ remaining after correction with $d$ and/or $d_3$, viz., $\Delta Z_{G=13} \equiv  0.8e_2\muas$. $c$ therefore can be interpreted as the average local offset of EDR3 parallaxes with respect to the rest of the sky/quasar frame of reference that the $\zf$ model was calibrated to. The pivot point of $\nueff = 1.48 \nueffunit$ is chosen to be consistent with the $q_{1k}$ and $q_{2k}$ color terms in the $\zf$ model, which take the form $(\nueff - 1.48)$ and $(1.48 - \nueff)^3$ \citep{l20b}. The pivot point of $G=10.8$ is motivated by the piecewise functions in magnitude used to describe the magnitude dependence of $\zf$, and which have a breakpoint at $10.8$. This is also in the transition region of $G \approx 11$ between window classes WC0a and WC0b. The faintest breakpoint in the $\zf$ model that overlaps with the sample's magnitude range occurs at $G=12.2$, which sets the pivot point for the $e$ term.

Note that one is able to simultaneously constrain 1) fractional errors in the asteroseismic radii, which would naturally arise from problems in the radius scaling relation (Eq.~\ref{eq:scaling3}) entering into the asteroseismic parallaxes (Eq.~\ref{eq:r_plx}), as parametrized by the $a$ and $d_a$ terms in Equation~\ref{eq:ast_corr}, as well as 2) additive errors in the corrected Gaia parallaxes, as parametrized by the $c$, $d$, and $e$ terms in Equation~\ref{eq:dz}.

The elements of the covariance matrix in Equation~\ref{eq:second_model} are given by
\begin{align}
\label{eq:inflation}
    C^{'}_{ij} &= C_{ij} + k^2\delta_{ij}\sigma^2_{\varpigaia} + \left[g^2 + \left(\frac{\partial Y}{\partial \varpiast}\right)^2 \right]\delta_{ij}\sigma^2_{\varpiast} +\nonumber\\ 
    &\left[\left(\frac{\partial Y}{\partial \nueff}\right)^2 + \left( \frac{\partial \Delta Z}{\partial \nueff} \right)^2\right]\delta_{ij}\sigma^2_{\nueff} + \left( \frac{\partial \Delta Z}{\partial G}\right)^2\delta_{ij}\sigma^2_{G} + \nonumber \\
    &\delta_{ij}s^2,
\end{align}
and describe the covariance between the asteroseismic-Gaia parallax difference for two stars, $i$ and $j$, where $\delta_{ij}$ is the Kronecker
delta function. In the above, $k$ and $g$ describe corrections to the
formal statistical uncertainties of Gaia and asteroseismic parallaxes,
which can be constrained since the two uncertainties are not strongly correlated (see \S\ref{sec:err}). $s$ can be thought of as an intrinsic scatter in the parallax residuals that capture any variation not described by the model, or, alternatively, as an additive correction to the Gaia and/or asteroseismic parallax uncertainties.  $\sigma_G$ is the uncertainty on $G$, and $\sigma_{\nueff}$, the uncertainty in \texttt{nu\_eff\_used\_in\_astrometry}, is adopted to be $1\%$ of $\nueff$, since the latter is a fixed and not fitted quantity for five-parameter astrometric solutions --- this is $\approx 2$ times larger than the uncertainty in \texttt{astrometric\_pseudo\_colour}, which is a fitted color as part of the six-parameter solutions. The statistical uncertainties in the asteroseismic parallaxes are denoted $\sigma_{\varpiast}$, and include contributions from $\dnu$, $\numax$, $\teff$, $A_{\ks}$, $\ks$, and $BC$ via linear propagation of uncertainty. The Gaia spatially-correlated errors are encapsulated in $C_{ij}$, which describe the spatial covariance in Gaia parallax between stars $i$ and $j$ separated by an angular distance, $\theta_{ij}$.

By appealing to quasars in EDR3, \cite{l20a} fit an exponential function to the parallax spatial covariance for $\theta_{ij} \gtrsim 0.5^{\circ}$ of $C_{ij\mathrm{,QSO}} = \rho_{\mathrm{QSO}} e^{\left ( -\ln 2 \theta_{ij}/\theta_{1/2,\mathrm{QSO}}
\right )}$, with $\rho_{\mathrm{QSO}} = 142 \muas^2$ and $\theta_{1/2,\mathrm{QSO}} \approx 11^{\circ}$. 

In this work, I derive an estimate of the spatial covariance matrix (\S\ref{sec:spatial}), modelling it with an equation of the same form as $C_{ij\mathrm{,QSO}}$:
\begin{equation}
\label{eq:spatial}
C_{ij} = \begin{cases}
      \rho  e^{\left (-\ln 2\theta_{ij}/\theta_{1/2}\right )} ,& \mathrm{for}\  0^{\circ} < \theta_{ij} \leq 10^{\circ} \\
      0,& \mathrm{otherwise},
    \end{cases}
\end{equation}
with best-fitting parameters according to Table~\ref{tab:spatial}. Note that $\theta_{1/2}$ is defined analogously to the half-angle in the Gaia team description of the covariance, $\theta_{1/2,\mathrm{QSO}}$. I tested the impact of spatial covariance in the likelihood analysis using an approximation described in \cite{zinn+2017plxtgas}. Briefly, I divided the Kepler field into $\sim 2.5\deg \times 2.5\deg$ squares corresponding to the Kepler modules, treating each one as independent from the other, given that spatially-correlated Gaia parallax errors are $\approx 1\%$ the level of statistical uncertainties on scales larger than $3\deg$ (see \S\ref{sec:spatial}). %STAT
 The results thus accounting for spatial correlations are not significantly different from those without spatial correlations, and so I neglect the spatial covariance term in Equation~\ref{eq:inflation}, $C_{ij}$, in what follows. I describe how I infer the spatial covariance in Gaia parallaxes and present the fit to Equation~\ref{eq:spatial} in \S\ref{sec:spatial}.

\section{Results and discussion}
\begin{figure}
    \centering
    \includegraphics[width=0.48\textwidth,trim=5cm 1cm 1cm 1cm,clip]{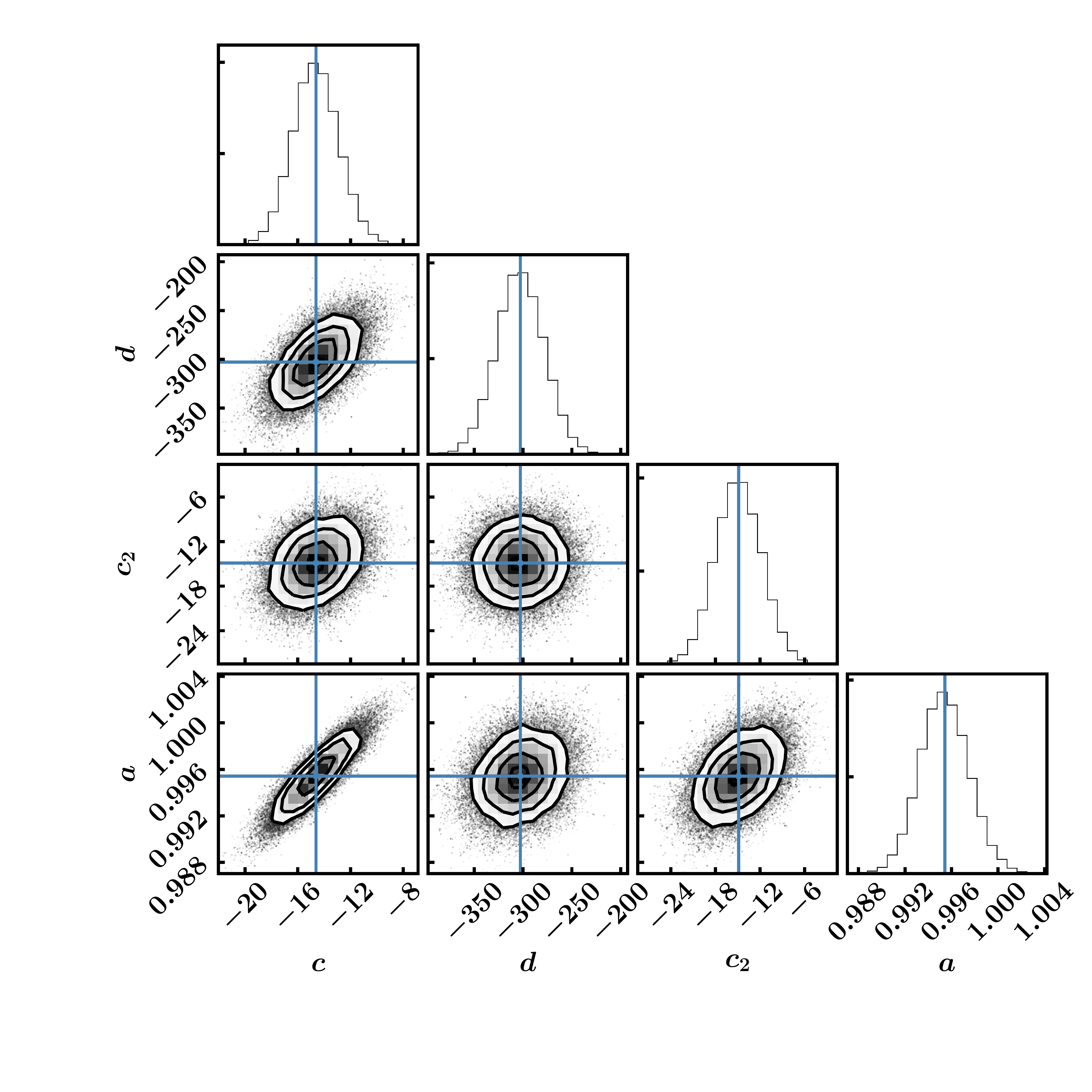}
    \caption{Posterior distributions of the parameters describing residual corrections to the $\zf$ model required to bring corrected Gaia parallaxes into agreement with asteroseismic parallaxes, according to Equations~\ref{eq:ast_corr}~\&~\ref{eq:dz}. Best-fitting parameter values are provided in Table~\ref{tab:res} (Model 0).}
    \label{fig:mcmc}
\end{figure}

I fit the model of Equation~\ref{eq:second_model} using MCMC, rejecting from analysis stars whose Gaia and asteroseismic parallaxes disagree by more than $2.5\sigma$. Figure~\ref{fig:mcmc} shows the resulting posterior distributions of the parameters for the preferred model (Model 0 in Table~\ref{tab:res}).  The best-fitting parameters are provided in Table~\ref{tab:res}, which are taken to be the means of the posteriors; the uncertainties are taken to be the standard deviations of the posteriors. I also provide the Bayesian Information Criterion difference  \citep[$\Delta$BIC;][]{schwarz1978} for all models that I considered compared to the preferred model, where a smaller value indicates a stronger evidence for the model, and a difference in 6 between models is taken to be strong preference \citep{kass_raftery1995}. In what follows, I take a conservative approach to potential adjustments to $\zf$, by default assuming terms in $\Delta Z$ are null unless there is strong evidence for them.

 Before correction by $\zf$, the raw Gaia and asteroseismic parallaxes have a mean difference of $+22\muas$ (scatter of $23\muas$), in the sense that Gaia parallaxes are smaller than asteroseismic parallaxes. Even before correction according to $\zf$, this is a significant improvement over the $\approx +50\muas$ under-estimation of DR2 parallaxes compared to asteroseismic parallaxes for this sample of red giant branch stars \citep{zinn+2019zp}.  For stars with $G > 10.8$, correcting the Gaia parallaxes according to $\zf$; adjusting the asteroseismic radii with $a$; and removing the local offset unique to the Kepler field, $c$, yields a residual of $+9\muas$, which is due to a color trend. I note that a non-zero color term, $d$, is strongly preferred to fit the parallax residuals. However, I take caution in interpreting this term as solely due to adjustments needed of $\zf$. As I explain in \S\ref{sec:color}, the color term as fitted in Model 0, $d$, may have contributions due to color-dependent asteroseismic parallax errors ($d_a$ in Eq.~\ref{eq:ast_corr}), for which there is not strong enough evidence to confirm. Under a conservative assumption, removing finally the color term implies the $\zf$ model leaves an insignificant residual of $+0.3\muas$, with an uncertainty in the mean of this residual of $0.4\muas$. For stars with $G \lesssim 11$, the $\zf$ model appears to over-correct the parallaxes by $-15 \pm 3\muas$, as I discuss in \S\ref{sec:mag}.

\begin{figure}
\centering
    \includegraphics[width=0.48\textwidth,trim=0cm 0cm 0cm 0cm,clip]{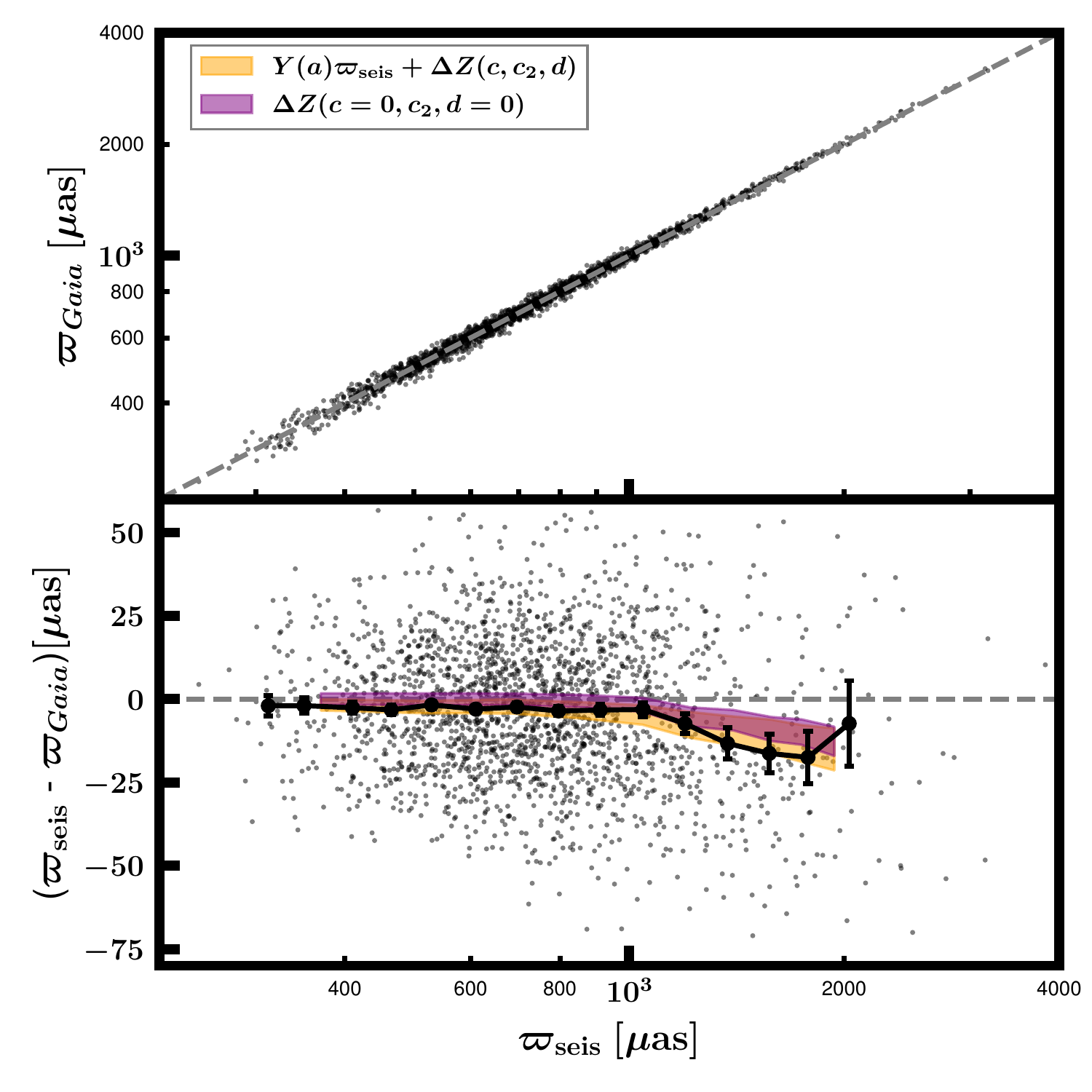}
    \caption{Residuals of the difference between asteroseismic and Gaia parallax corrected according to the \cite{l20b} $\zf$ model as a function of asteroseismic parallax, $\varpiast$. Black error bars show binned weighted means of the parallax residuals. The yellow band represents a running weighted mean of the best-fitting model (Model 0 in Table~\ref{tab:res}) described by Equations~\ref{eq:ast_corr}~\&~\ref{eq:dz}, and which includes an additive offset, a rescaling of the asteroseismic parallax, and a color-dependent term (the width of the band encompasses a $\pm 1\sigma$
  region in the offsets, $c$ and $c2$, and $\pm 0.5\sigma$ in the color term, $d$). The purple band shows a model without accounting for errors in the asteroseismic radii ($Y(a,d_a) = 0$; see Eq.~\ref{eq:ast_corr}, \S\ref{sec:color},~\&~\S\ref{sec:a}) or a local offset in Gaia parallaxes unique to the region of the sky analyzed here ($c=0$; see Eq.~\ref{eq:dz} and \S\ref{sec:off}).}
  \label{fig:vplx}
\end{figure}

\begin{figure*}
\centering
    \includegraphics[width=1.0\textwidth,trim=5cm 1cm 5cm 1cm,clip]{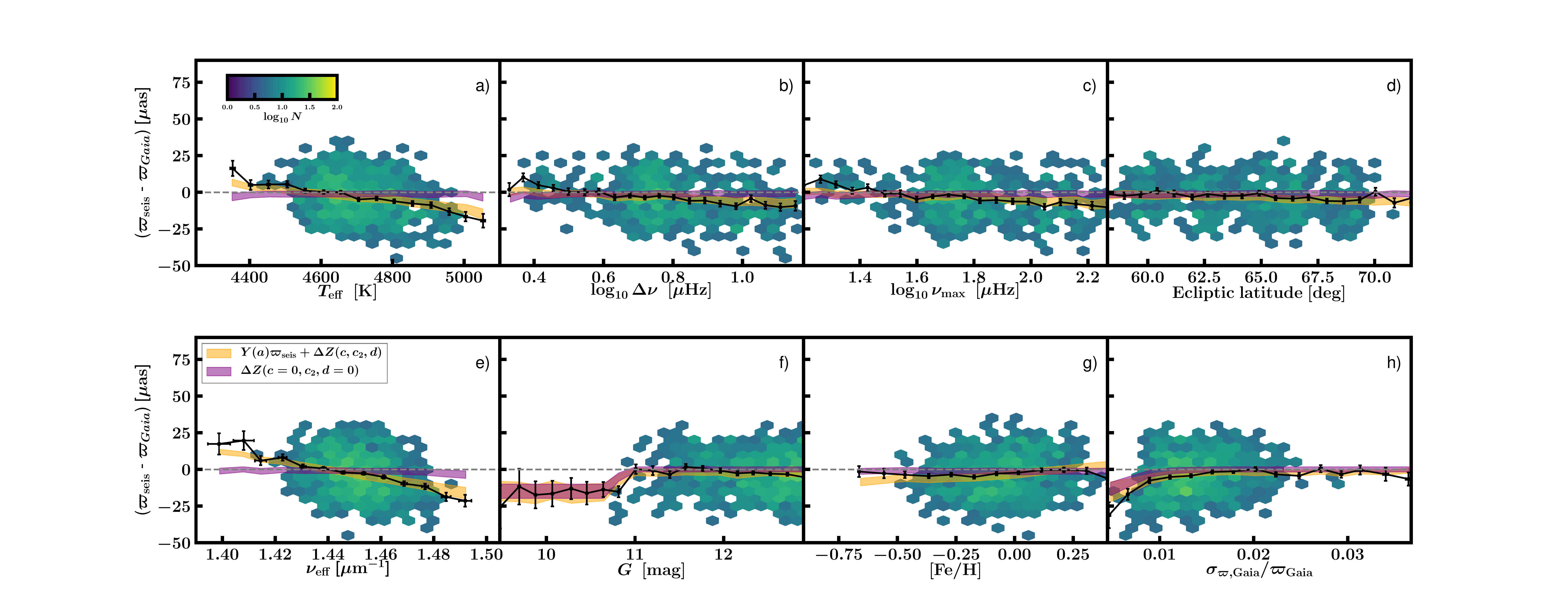}
    \caption{Residuals of the difference between asteroseismic and Gaia parallax corrected according to the \cite{l20b} $\zf$ model as a function of $\teff$ (a), $\dnu$ (b), $\numax$ (c),  ecliptic latitude (d),
  $\nueff$ (e), $G$ (f), $\mathrm{[Fe/H]}$ (g), and fractional Gaia parallax uncertainty, $\sigma_{\varpi\mathrm{,Gaia}}/\varpigaia$ (h). Black error bars show binned weighted means of the parallax residuals. The yellow band represents a running weighted mean of the best-fitting model (Model 0 in Table~\ref{tab:res}) described by Equations~\ref{eq:ast_corr}~\&~\ref{eq:dz}, and which includes an additive offset, a rescaling of the asteroseismic parallax, and a color-dependent term (the width of the band encompasses a $\pm 1\sigma$
  region in the offsets, $c$ and $c2$, and $\pm 0.5\sigma$ in the color term, $d$). The purple band shows a model without accounting for errors in the asteroseismic radii ($Y(a,d_a) = 0$; see Eq.~\ref{eq:ast_corr}, \S\ref{sec:color},~\&~\S\ref{sec:a}) or a local offset in Gaia parallaxes unique to the region of the sky analyzed here ($c=0$; see Eq.~\ref{eq:dz} and \S\ref{sec:off}), and which has the same meaning as the purple band in Fig.~\ref{fig:vplx}. The hexagonal bins represent the density of points, as indicated by the color bar in panel a; for clarity, bins with fewer than five points are not shown.}
  \label{fig:vall}
    \end{figure*}

The parallax residuals (asteroseismic -- $\zf$-corrected Gaia)
modelled here are shown as a function of asteroseismic parallax and various other parameters in Figures~\ref{fig:vplx}~\&~\ref{fig:vall}. The observed difference between corrected Gaia parallaxes and asteroseismic parallaxes are shown as error bars, and the best-fitting, preferred model to describe these residuals, $\Delta Z$ (Eq.~\ref{eq:dz} and parameters from Model 0 in Table~\ref{tab:res}), is shown as the yellow band. The purple band shows the model for what the parallax residuals would look like if the asteroseismic radii were exactly on the Gaia parallactic scale (i.e., $Y(a) = 0$), assuming there is no color term in $\Delta Z$ ($e=0$) or in asteroseismic parallax ($d_a = 0$), and in the absence of a local offset for the Kepler field's part of the sky ($c=0$). As such, it represents the $c_2$ term of Equation~\ref{eq:dz}, and which is the adjustment to the $\zf$ model for which there is the strongest evidence (\S\ref{sec:mag}).  The purple band thus demonstrates that the $\zf$ model performs very well for $G \gtrsim 11$, apart from a constant, local offset for the Kepler field not shown by the purple band ($c = -15 \pm 2\muas$ from Model 0 in Table~\ref{tab:res}). 

Below, I discuss aspects of the parallax residuals as a function of the different terms in Equation~\ref{eq:dz}, considering evidence for not only refinements to the $\zf$ model with a Kepler field--specific, local offset, $c$, a color term, $d$, and a magnitude term, $e$, but also evidence for corrections to the asteroseismic parallaxes with the color term, $d_a$, and the asteroseismic radius rescaling term, $a$. I also discuss the uncertainty budget in the parallax difference, including evidence for corrections to the fractional parallax uncertainties and quantifying  systematic spatial variations in Gaia parallaxes.

\begin{figure}
    \centering
    \includegraphics[width=0.48\textwidth,trim=0cm 0cm 0cm 0cm,clip]{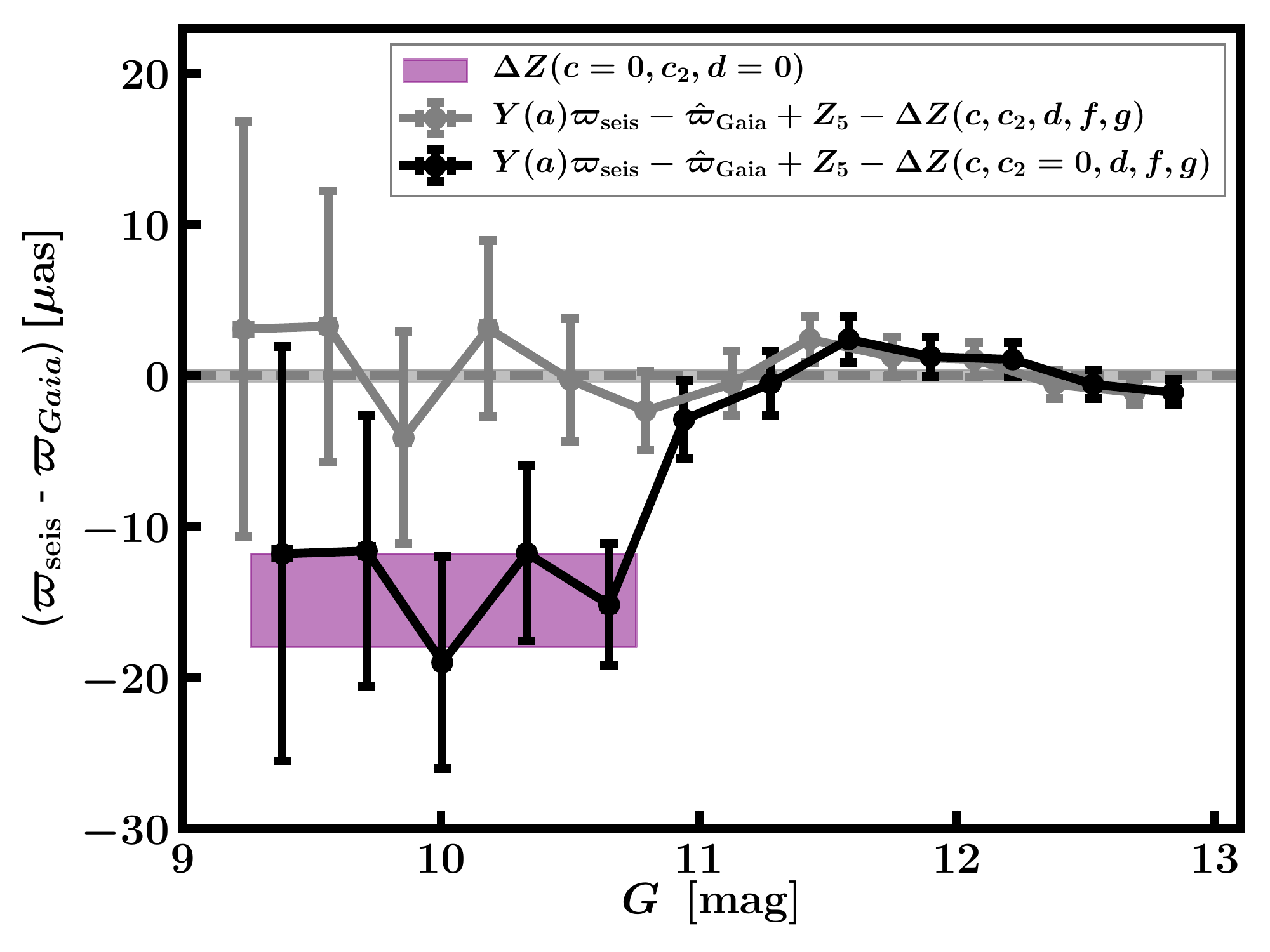}
    \caption{Binned, weighted means of the difference between
      asteroseismic and Gaia parallax corrected according to the
      \cite{l20b} $\zf$ model as a function of $G$ are shown as black
      error bars, after further corrections with a color term, a
      constant offset, and a rescaling of the asteroseismic radii
      (according to Equations~\ref{eq:ast_corr}~\&~\ref{eq:dz} with
      parameters from Model 0 in Table~\ref{tab:res}, except with
      $c_2=0$). The purple shaded region represents the additive
      adjustment, $c_2$, to the Gaia team's $\zf$ parallax zero-point model
      required to bring $\zf$-corrected EDR3 parallaxes into alignment
      with asteroseismic parallaxes (the band width corresponds to the
      $1\sigma$ confidence interval for $c_2$); corrected EDR3
      parallaxes for $G < 10.8$ are thus inferred to be $15\pm 3\muas$
      too large. Other terms in Equation~\ref{eq:dz} are either not
      strongly preferred by the data or likely to be specific to the
      Kepler field (e.g., $c$). After the adjustment of the $\zf$
      model according to the purple band, the parallax residuals
      reduce to the grey error bars, which have been shifted
      horizontally for clarity. The grey band indicates that the
      $\zf$-corrected Gaia parallaxes agree with asteroseismic
      parallaxes for stars in the sample with $9 \lesssim G \lesssim
      13$, $1.4 \nueffunit \lesssim \nueff \lesssim 1.5 \nueffunit$ to
      within on average $\pm0.4\muas$ after the bright-end adjustment
      to $\zf$ of $c_2 = -15\muas$.}
    \label{fig:res}
\end{figure}

\subsection{Corrections to Gaia and asteroseismic parallaxes: the color term}
\label{sec:color}
It should first be noted that the color term in this analysis, $d$, describes a smaller range in $\nueff$ ($1.4 \nueffunit \lesssim \nueff \lesssim 1.5\nueffunit$) than the linear color terms $q_{1k}$ of the $\zf$ model attempt to explain ($1.24 \nueffunit \lesssim \nueff \lesssim 1.72\nueffunit$). This means that stronger local gradients in the parallax systematics may still remain after correction by $\zf$, which the sample could be sensitive to. The sample will be most sensitive to color trends between $12 \lesssim G \lesssim 13$, where most of the data are, meaning that any color-dependent refinement to $\zf$ I find are not necessarily valid in other magnitude regimes, and may not describe color trends within the brighter range of the sample $9 \lesssim G \lesssim 12$.

With these caveats in mind, a significant color term of $d = -300 \pm 25 \dunit$ is strongly favored to describe the parallax residuals (Model 11 vs. Model 0). This trend is also clearly visible in Figure~\ref{fig:res}e. 

Here, I take care to consider the possibility that $d$ may not only describe an additive correction to EDR3 parallaxes, but also may partially describe small problems in the asteroseismic parallax via the bolometric correction and/or $\dnu$ correction.

Regarding possible contributions to $d$ from the asteroseismic parallaxes, consider, for example, that bolometric corrections used to calculate the asteroseismic parallax (Eq.~\ref{eq:r_plx}) are found to vary as a function of temperature/color depending on the prescription by $\approx 2-4\%$ \citep{zinn+2019rad,tayar+2020}. This is enough to explain some of the color dependence of the parallax difference: for a median parallax of the sample of $\approx 0.7\mas$, the resulting effect on the asteroseismic parallaxes would be $7-14\muas$, corresponding to differences in $\approx 70-100\mathrm{\dunit}$ in $d$, which is comparable to the best-fitting color term of $d_2 = -300 \pm 25 \dunit$. 

Additional color residuals may come about through the choice of the $\dnu$ correction, which depends on temperature and therefore color. I test the sensitivity of the color trend to $\dnu$ corrections by considering the non-linear scaling relations of \cite{kallinger+2018} instead of the APOKASC-2 $\dnu$ corrections (Model 16). The resulting color term of $-210 \pm 25\dunit$ reveals a $\approx 100 \dunit$ level variation due to $\dnu$ correction choice. Similarly, the \cite{yu+2018} data using the same non-linear scaling relations yield $d = -180 \pm 30\dunit$ (Model 14). Taken together, these differences of $\sim 100 \dunit$ are suggestive of the level of color-dependent uncertainties in $\dnu$ corrections used to calibrate asteroseismic radii. 

Due to these indications of contributions to $d$ from asteroseismic parallax systematics instead of EDR3 parallax systematics, I considered a color-dependent correction explicitly for the \textit{asteroseismic} parallaxes of the form $d_a \varpiast(\nueff - 1.48)$ (Eq.~\ref{eq:ast_corr}). This term is designed to capture color systematics in the asteroseismic parallax, which will tend to be fractional (dependent on $\varpiast$), given that bolometric corrections and $\dnu$ enter multiplicatively in Equation~\ref{eq:r_plx}. (The same rationale motivates the fractional correction to asteroseismic parallaxes, $a$, in Equation~\ref{eq:ast_corr}.) Simultaneously allowing for an additive color term, $d$, and this fractional term, $d_a$, yields $d_a=-0.14\pm 0.06 \daunit$ and a less substantial $d = -220\pm 40\dunit$  (Model 2). For the typical star in the sample with $\varpiast \approx 700\muas$, this value of $d_a$ corresponds to a color term due to systematics in the asteroseismic parallax of $\approx -100\dunit$, which is consistent with expectations as outlined above. Nevertheless, there is not strong evidence (i.e., $\Delta$BIC $< -6$) for adding $d_a$ either as a replacement for the additive color correction (Model 9 vs. Model 0) or in addition to $d$ (Model 2 vs. Model 0). I therefore conservatively correct for a color term in what follows, without definitely attributing it fully to Gaia parallax systematics. To be sure, a color correction of some sort is required to explain the parallax residuals, though it is likely partially due to an additive adjustment needed of $\zf$ as well as a properly parallax-dependent color term arising from bolometric correction and $\dnu$ systematics.

Non-Gaia contributions notwithstanding, there would appear to remain a $\approx -200\dunit$ residual that would be attributable to color residuals in the $\zf$ model. In this regard, one can helpfully refer to the consistency check performed by the Gaia team in the LMC. On average, there is not a large gradient in the LMC parallax across the whole range $1.1 \nueffunit \lesssim \nueff \lesssim 1.9\nueffunit$ after correction by the Gaia zero-point model (Fig. 23 of \citealt{fabricius+2020}). However, it is clear that gradients exist on smaller scales, which, for $1.4 \nueffunit \lesssim \nueff \lesssim 1.5\nueffunit$, is approximately $-140\dunit$, in the sense that bluer stars have too-small parallaxes. Although most of the LMC members used by the Gaia team have $G > 13$, this nonetheless indicates that linear residuals in parallax as a function of color exist after correction by $\zf$. Moreover, this gradient is similar in magnitude but opposite in sign to the gradient I identify. Recalling that the sine of the ecliptic latitude of the LMC and Kepler field are approximately the same magnitude ($\approx 1$) but of opposite sign, this is suggestive of a refinement to the $q_{01}$ term for $G=12.2$ in the $\zf$ model: $q_{01}$ sets the magnitude of a correction of the form $\sin \beta (\nueff - 1.48)$, and an increase from $\approx 40\dunit$ adopted for the $\zf$ model to at least $100\dunit$ could seemingly be accommodated, given the uncertainties in $q_{01}$ (Fig. 11 of \cite{l20b}). Keeping in mind both the potential errors in the $\zf$ coefficients due to the bootstrap fitting approach in the bright regime ($G < 10.8$; \citealt{l20b}) and the two times larger range in color explained by the $q_{01}$ term as opposed to $d$, I believe this is a plausible scenario.

I attempted to model the color-dependent residuals with the addition of a cubic term in the color dependence ($d_3$ in Eq.~\ref{eq:dz}), which was motivated by the Gaia team's cubic term in $\zf$, $q_{20}$. However, $d_3$ is strongly disfavored compared to having no cubic term, according to the $\Delta$BIC in Table~\ref{tab:res} (Model 5 vs. Model 0). The data also prefer having a single color term instead of one to describe $ G < 10.8$ and one to describe $G \geq 10.8$, as is seen from the difference in $\Delta$BIC between Model 4 and Model 0.

\subsection{Corrections to asteroseismic parallaxes: the radius rescaling term}
\label{sec:a}
I note that the radius rescaling term of $a = 0.995 \pm  0.002$ %STAT
for the preferred model (Model 0 in Table~\ref{tab:res}) is consistent within $2\sigma$ with the value of $1.015 \pm 0.003 \,\mathrm{(stat.)} \pm 0.013\,\mathrm{(syst.)}$ for stars with $3.5\rsun \leq R \leq 10\rsun$ as well as the value of $1.019 \pm 0.006\,\mathrm{(stat.)} \pm 0.013\,\mathrm{(syst.)}$ for $10\rsun < R < 30\rsun$ from \cite{zinn+2019rad} in an analysis using the APOKASC-2 first-ascent red giant sample and Gaia DR2 data. The systematic uncertainty in this case is not the full systematic uncertainty from that analysis, but rather just the systematic uncertainty arising from the Gaia DR2 zero-point ($1.3\%$), since otherwise the same asteroseismic parallax data were used (apart from small differences in the APOGEE DR14 and DR16 temperatures). 

The shifts of up to $0.021$ in $a$ for Model 14 and Model 16 compared to Model 0 are to be expected, since they both use the \cite{kallinger+2018} non-linear scaling relations (Eq.~\ref{eq:scaling3}), which may differ systematically from the APOKASC-2 radius scale described by Equation~\ref{eq:scalingapo}. Although the \cite{kallinger+2018} non-linear scaling relations are calibrated to some of the same data used by the \cite{pinsonneault+2018} calibration, the methodologies differ. Indeed, one might expect a $1\sigma$ systematic difference in $a$ of $0.018$, which includes the variation in the Gaia DR2 zero-point \citep{zinn+2019zp}; a $0.7\%$ variation in the APOKASC-2 radius scale due to the calibration to open cluster masses \citep{pinsonneault+2018}; and an uncertainty of $1.1\%$ due to possible variation in the non-linear exponents, $\kappa$ and $\gamma$, of Equation~\ref{eq:scaling3}.

\subsection{Corrections to Gaia parallaxes: the constant offset term}
\label{sec:off}
I consider the local offset of $c=-15\pm 2\muas$ to be a conservative estimate of the mean deviation of parallaxes in the Kepler field from the parallax scale of quasars to which the $\zf$ model is tied. This is because the local offset is defined, consistent with the Gaia team's $\zf$ model, at $\nueff = 1.48\nueffunit$, and so differences in the color trend in the parallax residuals would tend to shift $c$. For a fixed color trend in the residuals, the precise value of $c$ will also depend on choice of the pivot point, again, taken here to be $\nueff = 1.48\nueffunit$. Nevertheless, I do consistently find evidence across models for a local offset, $c$, of $\approx -15\muas$, even for Models 14-17, which prefer a $\approx 100\dunit$ less substantial color term than that of Model 0. 

Perhaps more importantly, Models 14-17 all use the non-linear scaling relations (Eq.~\ref{eq:scaling3}) instead of the APOKASC-2 $f_{\dnu}$ (Eq.~\ref{eq:scalingapo}), and prefer a different, more significant radius rescaling, $a$ (\S\ref{sec:a}). This suggests the fractional and additive corrections to parallaxes described by Equations~\ref{eq:ast_corr}~\&~\ref{eq:dz} are well-fit, even as there are correlations in the posteriors of $c$ and $a$ (Figure~\ref{fig:mcmc}). 

In spite of this conservative estimate, the best-fitting $c$ is broadly consistent with the $8.1\muas$ root-mean-square (RMS) variation expected of Gaia parallaxes due to spatial correlations on scales the size of the Kepler field and larger inferred by the Kepler team using quasars (\citealt{l20a}). For the analysis of spatial correlations on scales smaller than $10\deg$, see \S\ref{sec:spatial}.

\subsection{Corrections to Gaia parallaxes: the magnitude term}
\label{sec:mag}
As can be seen in Figure~\ref{fig:vall}f, a linear magnitude dependence is observed in the parallax residuals for $11 \lesssim G < 13$. 
Attempting to explain this trend with an explicit magnitude-dependent correction to the Gaia parallaxes is not strongly favored by the data: neither a model with the addition of a magnitude term (Model 3), nor a model with a magnitude term instead of a radius rescaling term (Model 1) is strongly preferred over a model without a magnitude term (Model 0). Rather, this trend is well-described by the radius rescaling factor, $a$ (\S~\ref{sec:a}), without the need for an explicit magnitude term, $e$. This is because the magnitude of a giant is correlated with its parallax, and residuals between asteroseismic and Gaia parallax due to a fractional asteroseismic radius error will algebraically tend to increase in magnitude with increasing parallax.

As I do in interpretations of color-dependent adjustments to $\zf$ in \S\ref{sec:color}, I adopt a conservative assumption that the $\zf$ model does not require adjustments in the absence of strong evidence. The preferred model therefore does not include $e$. Two other lines of evidence from independent work suggest that there is no need for refinements to $\zf$ that are linear in magnitude: 1) there is no evidence for a magnitude term from independent tests using Cepheids \citep{riess+2020}, and 2) the \cite{yu+2018} data prefer a magnitude term consistent with zero ($e = 0.0 \pm 1.1 \muas\,\mathrm{mag^{-1}}$; Model 15). In short, I believe that the linear magnitude dependence in the parallax residuals of Figure~\ref{fig:vall}f should be thought of instead as an error in the asteroseismic radius of $\approx 0.5\%$, which is the required level to explain the trend with magnitude in the absence of a magnitude term (Model 0).

It should also be noted that whatever refinements may be required of
$\zf$ may not be captured by the model, since I assume a linear
dependence in magnitude across a range of $\sim 4$ magnitudes; an
attempt to make the magnitude trends of Equation~\ref{eq:dz} more
granular by fitting two separate terms for $G< 10.8$ and $G \geq 10.8$
is strongly disfavored (Model 7 vs. Model 0). By contrast, the $\zf$
model is more fine-grained, describing magnitude-dependent systematics
with piece-wise functions with four breakpoints ($G={10.8, 11.2, 11.8,
  12.2}$) within the magnitude range of the sample \citep{l20b}. (The
$\Delta Z$ model in this analysis, however, does provide a more local measurement of color systematics than the $\zf$ model [\S\ref{sec:color}].)

Apart from the small trend with magnitude seen in Figure~\ref{fig:vall}f, there is an abrupt shift in the parallax residuals at $G \approx 11$. There is no plausible astrophysical reason for why there should be a discontinuous parallax difference at $G \approx 11$, and is instead indicative of shortcomings in the Gaia $\zf$ model, given that 1) this is the magnitude at which there is a window transition between WC0a  and WC0b \citep{l20a}, and that 2) $G=10.8$ is also a breakpoint in the magnitude-dependent terms of the $\zf$ model (thus motivating the choice for pivot point in Equation~\ref{eq:dz}). I model this with the $c_2$ term, which is effectively a local offset just for the brightest stars in the sample ($G < 10.8$). I find strong evidence for a non-zero value of $c_2 = -15 \pm 3\muas$ (Model 8 vs. Model 0), in the sense that corrected Gaia parallaxes for $G < 10.8$ are too large.

EDR3 parallaxes are also found to be too large by $-14 \pm 6\muas$
compared to photometric parallaxes from classical Cepheids with colors
$1.35\nueffunit < \nueff < 1.6 \nueffunit$ and brighter than $G
\approx 11$ \citep{riess+2020}. \cite{bhardwaj+2020} find a similar
offset of $-25 \pm 5\muas$ compared to RR Lyrae ($ \nueff >
1.5\nueffunit$, $G \lesssim 12$). $c_2$ is also consistent to within
$2\sigma$ with the offset of $+15 \pm 18\muas$ from eclipsing binaries
($5 \lesssim G \lesssim 12$, $\nueff > 1.5\nueffunit$;
\citealt{stassun_torres2021}). More recently, an analysis using
photometric red clump parallaxes indicates an offset for $G < 10.8$
with a magnitude of $9.8\pm 1\muas$ \citep{huang+2021}. An effect of $\approx +10\muas$ in the opposite direction was noted when validating the $\zf$ model with LMC stars ($G < 13$, $\nueff > 1.7\nueffunit$) \citep{l20b}. Comparisons using the Washington Double Star Catalog \citep{mason+2001} also showed a $+10\muas$ offset in the opposite direction between the data and the $\zf$ model for stars $G < 7$ \citep{fabricius+2020}. 

 One possible explanation of the bright-end parallax systematic I find here is that the $\zf$ model does not have the correct $q_{00}$ coefficients, which are weights in the $\zf$ model that describe constant, additive corrections to EDR3 parallaxes at particular magnitude breakpoints, with linear ramps between adjacent breakpoints. The required increase in $q_{00}$ of $\approx 15 \muas$ appears inconsistent with the statistical uncertainties for $q_{00}$ coefficients for breakpoints $G={6.0, 10.8, 11.2}$ that contribute to the $\zf$ model for the $10 < G < 10.8$ range (Fig. 11 of \cite{l20b}). However, it should be noted the bright-regime solution for $\zf$ is iterative: depending on 1) an intermediate solution bridging the quasar+LMC $\zf$ solution valid for $G > 13$ and one that includes bright wide binaries down to $ G = 10$ and 2) a subsequent iterative solution that includes additional wide binaries with $ G < 6$. This could plausibly cause systematic errors in the bootstrapped solution for $G < 10.8$, as noted by \cite{l20b}. Although an increase in $q_{00}$ for $ G \lesssim 11$ would worsen the too-small blue LMC stars with $G \lesssim 13$ noted by \cite{l20b}, there are too few stars with $ G \lesssim 11$ in Figure 19 of \cite{l20b} to discern if the too-small problem actually exists for $G \lesssim 11$ rather than only $11 \lesssim G \lesssim 13$. It is also plausible that the too-blue LMC stars are subject to their own systematics altogether unrelated to and unaffected by any $c_2$ or indeed any other systematic identifiable with the sample: the too-small parallaxes occur for stars that have $\nueff > 1.72\nueffunit$, which is color regime with no chromaticity calibration for five-parameter solutions and which is a regime not fitted with a linear color term in $\zf$, but rather a constant of $0.24 \times q_{10}$ \citep{l20b}.

The best-fitting $c_2$ term is shown in Figure~\ref{fig:res}, where
black error bars indicate the parallax residuals after correction of
the asteroseismic parallaxes by the radius rescaling factor, $a$, and
after removing residual color trends, and removing a local offset with
the $c$ term, according to Equation~\ref{eq:dz} and the parameters of
Model 0 in Table~\ref{tab:res}. The purple shaded region therefore
represents the bright-end $c_2$ adjustment to the $\zf$ model, which
is the term for which I find the strongest evidence among all the
terms in Equation~\ref{eq:dz}. (The purple band in
Figure~\ref{fig:res} corresponds to the purple band in
Figures~\ref{fig:vplx}~\&~\ref{fig:vall}.) Because this analysis is limited to a single $10\times10$ sq. deg. area in sky coverage, it is conceivable that the $c_2$ term could be specific to the ecliptic latitude of the Kepler field. However, given its concordance with a very similar term found using Cepheids distributed across the sky \cite{riess+2020} and red clump stars distributed across the sky \citep{huang+2021}, I believe that $c_2$ is likely to be universal for $G < 10.8$.

\subsection{Comparison to Gaia DR2}
As noted above, the difference between the asteroseismic parallaxes and the raw Gaia EDR3 parallaxes without applying the $\zf$ correction is $+22\muas$ in the sense that the EDR3 parallaxes are smaller. This is a significant improvement from the Gaia DR2 global zero-point inferred for this red giant sample in \cite{zinn+2019zp} of $+53\muas$. Some combination of the improved basic angle modelling with VBAC and photometric modelling incorporated into the astrometric solution procedure and the longer observation period in EDR3 compared to DR2 contributes to this systematics reduction.

Regarding color- and magnitude-dependent terms in EDR3 compared to
DR2, the $e = -4.2 \pm 0.8 \muas\,\mathrm{mag}^{-1}$ magnitude term
found in \cite{zinn+2019zp} is in the same direction and of similar
magnitude to the value of $-2.6 \pm 0.9 \muas\,\mathrm{mag}^{-1}$ we
find here for Model 1. Nevertheless, a nonzero $e$ is not preferred by the EDR3 data (see \S\ref{sec:mag}). A color term, on the other hand, is significantly preferred by the EDR3 data, which I find to be larger in magnitude than the color term of $d = -220 \pm 21 \dunit$ found by \cite{zinn+2019zp} in Gaia DR2 parallaxes when using the same $\ks$-band bolometric correction adopted here. This difference may be partially caused by the different definition of $\nueff$ in this work and \cite{zinn+2019zp}: in DR2, the data model included \texttt{astrometric\_pseudocolour}, which was computed and defined analogously to the \texttt{pseudocolour} quantity in the EDR3 data model, which is now only provided for the six-parameter solutions. Indeed, adopting the DR2 \texttt{astrometric\_pseudocolour} values as $\nueff$ reduces the inferred $d$ for EDR3 to $\approx -270 \pm 20 \dunit$, which is consistent within $2\sigma$ with DR2. That the EDR3 and DR2 color terms are comparable suggests that neither the improvements in EDR3 to chromaticity corrections in the image parameter determination \citep{rowell+2020} nor the $\zf$ model completely correct color-dependent parallax systematics.

\subsection{Statistical uncertainty in Gaia parallaxes}
\label{sec:err}
In spite of nominally more than doubling the parallax precision for $9 \lesssim G \lesssim 12$ in EDR3 compared to DR2 \citep{l20a}, there are indications that EDR3 statistical parallax uncertainties are increasingly under-estimated for brighter stars --- \cite{l20b} and \cite{fabricius+2020} note that the uncertainties appear to be under-estimated by at least $30\%$ in the $G < 13$ regime. In an independent analysis using wide binaries, \cite{elbadry} estimate that sources with no close companions of similar brightness have EDR3 parallax uncertainties under-estimated by $20-30\%$ for $G < 13$.

For the fiducial analysis, I did not modify the nominal Gaia parallax or asteroseismic parallax uncertainties, setting $k=g=1$ (Model 0). Given that the $\chi^2/\mathrm{dof}$ for Model 0 is significantly smaller than unity (Table~\ref{tab:res}), I consider in Model 12 what $k$ (rescaling of the Gaia parallax) and $g$ (rescaling of the asteroseismic parallax) would be preferred by the model (Equations~\ref{eq:second_model}~\&~\ref{eq:inflation}). I infer that the Gaia parallax uncertainties are too small by $22\pm 6\%$, and also that asteroseismic parallax uncertainties are too large by $31\pm3\%$.

 Regarding the inferred over-estimation of the asteroseismic parallaxes, I note that the APOKASC-2 asteroseismic uncertainties are estimated conservatively \citep{pinsonneault+2018}. The uncertainties were taken to be the observed scatter in the pipeline results for each star, imposing a lower bound of $0.9\%$ and $0.4\%$ in $\numax$ and $\dnu$, respectively. The rationale behind this approach is well-motivated, i.e., so as to not allow unreasonably small uncertainties due to chance agreements among the pipeline measurements, but does impose an artificial noise floor. The resulting statistical uncertainty estimates were seen to result in too-low chi-squared per degrees of freedom in the mean mass of red giants in NGC 6791 and NGC 6819 of 0.6 and 0.8, confirming the conservative nature of the uncertainty estimates.
  The required deflation factor would leave uncertainties in $\dnu$ and $\numax$ still larger than the lower bound estimates of the intrinsic scatter in the $\dnu$ and $\numax$ parts of red clump asteroseismic scaling relations of $0.1 \pm 0.2\%$ for $\dnu$ and $0.7 \pm 0.2\%$ for $\numax$ \citep{li+2020}.

Concerning the level of EDR3 parallax uncertainty under-estimation, I
confirm results of \cite{fabricius+2020} and \cite{elbadry}, finding
an average under-estimation of $22 \pm 6\%$ in the Gaia parallaxes in
the magnitude range $9 \lesssim G \lesssim 13$ probed by the sample. I
note that the under-estimation effect appears to depend on magnitude
\citep{fabricius+2020,elbadry}, suggesting that $k$ varies within the
sample, which spans a range of approximately four magnitudes. I
therefore corrected the Gaia parallax uncertainties according to the
magnitude-dependent function of \cite{elbadry}, to infer if there were
a need for further correction to the Gaia parallaxes. The resulting
$g$ (Model 12 in Table~\ref{tab:res}) is consistent with unity, which
indicates both that 1) the approach in this analysis is sensitive to estimating both asteroseismic and Gaia parallax uncertainty corrections (a $30\%$ inflation of the asteroseismic parallaxes is still inferred in this case, even as $g=1.02 \pm 0.04$) and that 2) the \cite{elbadry} corrections perform well.

One can also compare how much of the Gaia uncertainty under-estimation may be due to the spatial covariance estimated in \S\ref{sec:spatial}. I estimate that the uncertainty in the mean Gaia parallax in the sample when including the spatial covariance of Equation~\ref{eq:spatial} and Table~\ref{tab:spatial} is $15\%$ larger than the uncertainty from the reported EDR3 statistical uncertainties, which implies that $\sim 1/2$ of the under-estimation of the statistical uncertainties may be due to spatial covariance induced notionally by the scanning pattern of the Gaia satellite. As noted by both \cite{fabricius+2020} and \cite{elbadry}, there is a tendency for crowded regions to have more significantly under-estimated parallax uncertainties than less crowded regions, which may be a contributing factor in addition to the variance induced by spatial correlations.

An additive correction to the Gaia parallax uncertainties --- which could also equally be interpreted as an intrinsic scatter in the asteroseismic-Gaia parallax difference unexplained by the rest of the model --- is not favored ($s$ in Equation~\ref{eq:inflation}; Model 6 vs. Model 0). In other words, the Gaia parallax uncertainties are well-described by being fractionally under-estimated rather than requiring additive corrections.

\subsection{Spatial correlations in Gaia parallaxes}
\label{sec:spatial}
\begin{figure}
    \centering
    \includegraphics[width=0.48\textwidth]{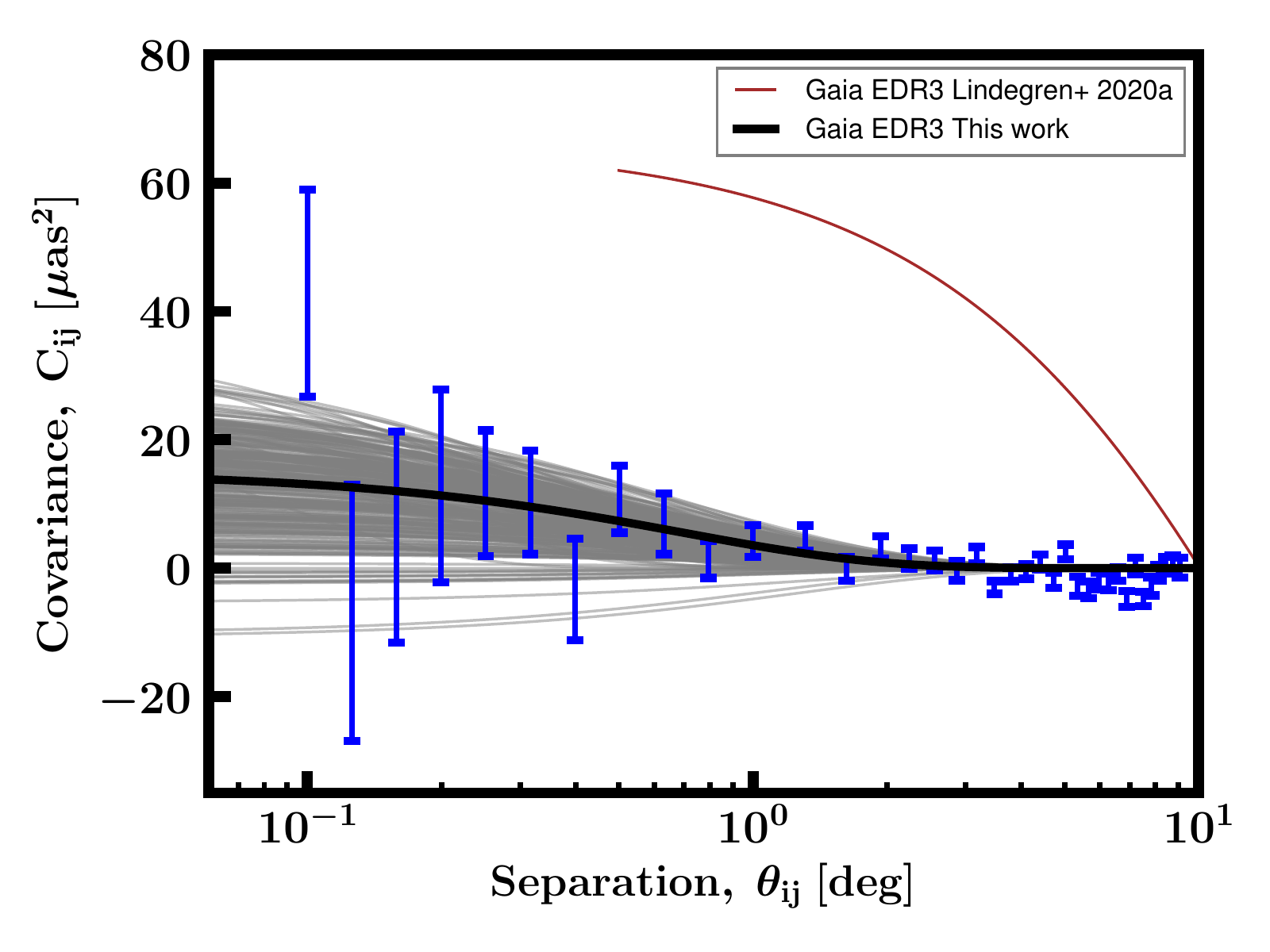}
    \caption{Parallax covariance for scales probed by the Kepler field ($0.1 \deg \lesssim \theta_{ij} \lesssim 10\deg$), as inferred from the differences between asteroseismic and Gaia parallaxes after correction according to $\zf$ and Equations~\ref{eq:ast_corr}~\&~\ref{eq:dz} (blue error bars). The best-fitting model of the form Equation~\ref{eq:spatial} with parameters according to Table~\ref{tab:spatial} is shown as the black curve. Random draws for the best-fitting model parameters consistent with the covariance among the parameters are shown as grey curves. For comparison, a model of quasar parallax covariance from \cite{l20a} found for fainter magnitudes ($G > 17$) and for larger angular scales ($\theta_{ij} \gtrsim 0.5\deg$) is shown as the brown curve, less the expected large-scale covariance. See \S\ref{sec:spatial} for details.}
    \label{fig:spatial}
\end{figure}
In addition to pointing out under-estimation of statistical
uncertainties in EDR3 parallaxes, the Gaia team quantifies systematic
variations in parallax as a function of position on the sky induced by
the scanning pattern of the satellite \cite{l20a}. The present
analysis extends the Gaia team's analysis to smaller scales, redder
colors, and brighter magnitudes compared to their estimate with faint,
blue quasars. The operating principle behind the estimate provided
here is that asteroseismic parallaxes are astrophysically expected to have no intrinsic correlation with position in the $10\times10$ sq. deg. Kepler field and thus any spatial systematics in EDR3 parallaxes would manifest as spatially-correlated differences between asteroseismic and EDR3 parallaxes.

In detail, the estimate of the spatial correlations in Gaia parallaxes follows \cite{zinn+2017plxtgas}.\footnote{Following \cite{zinn+2019rad}, I do not consider the uncertainty in the estimation due to `cosmic variance': the finite sampling of the spatial correlations due to looking only at the Kepler field. I do, however, take into account edge effects through bootstrap sampling, according to \cite{zinn+2017plxtgas}.} I correct the Gaia parallaxes according to the $\zf$ prescription and also according to Model 0, further subtracting off any residuals by forcing the Gaia parallaxes to have the same mean parallax as the asteroseismic parallaxes.  I then compute the covariance of the parallax difference $\varpiast - \varpigaia$, $C_{ij} = \langle (\varpiast - \varpigaia)_i(\varpiast - \varpigaia)_j \rangle $ for pairs of stars, $i$ and $j$, in bins of angular separation, $\theta_{ij}$, estimating the uncertainties in the covariance using from bootstrap sampling \citep{zinn+2017plxtgas}.\footnote{Whereas \cite{zinn+2017plxtgas} computed a binned Pearson correlation coefficient, I compute here the covariance, which differs by a factor of the variance in the parallax difference.} 

The resulting covariance at given angular scales are shown as error bars in Figure~\ref{fig:spatial}. A positive value indicates a correlation between parallaxes at a given angular separation, and a negative value indicates an anti-correlation. If there were no spatial correlations in the Gaia parallaxes, the covariance would be zero.

I then fit the covariance with the model of Equation~\ref{eq:spatial} via MCMC, adopting uncertainties on each binned covariance point as shown in Figure~\ref{fig:spatial}. The resulting model is shown as a solid black curve in Figure~\ref{fig:spatial}. The grey curves are random draws of the best-fitting model, according to the covariance among the best-fitting parameters, as estimated from the MCMC chains.

I show for reference the Gaia team covariance model from \cite{l20a} as the solid brown curve. I note that the Gaia model is fit to large scales $\theta_{ij} \gtrsim 0.5\deg$, and so I only show the \cite{l20a} model in that regime. I have also subtracted the variance inferred on scales larger than the Kepler field of $76\muas^2$ \citep{l20a} %STAT2 
to make it more comparable to the results of this analysis, though I expect residual differences to remain (see below).

As can be seen in Figure~\ref{fig:spatial} and from the best-fitting parameters in Table~\ref{tab:spatial}, the systematic covariance floor at the smallest scales is $4.0 \pm 1\muas$,%STAT3
which corresponds to $35 \pm 10 \%$ %STAT3
of the statistical uncertainty of $\varpigaia$ in the sample. Already at a low level on small scales, the covariance becomes negligible on scales larger than a few degrees.

In the overlap region of $0.5\deg \lesssim \theta_{ij} \lesssim
10\deg$ between the covariance estimated here and that estimated by
\cite{l20a} using quasars, the quasar covariance model is markedly
larger than the covariances of the Kepler field (brown versus black
curves in Fig.~\ref{fig:spatial}). This may be indicative of smaller
spatial systematics in the $G < 13$ regime compared to the $G > 17$
regime probed by quasars, or may signal variation in the small-scale
systematics across the sky. Note that there will be an offset between
the covariance estimated here using the Kepler field and the covariance inferred from quasars, since the latter has contributions from scales larger than $10\deg$. As noted above, I attempt to correct for this difference by removing $8.7\muas$ in quadrature from the quasar covariance, which is the estimated RMS scatter in quasar parallaxes on scales larger than $7\deg$ \citep{l20a}; a large-scale RMS scatter of $\approx 11\muas$ would be required to reconcile the quasar covariance and the Kepler field covariance at $0.5\deg$, which is also consistent with the Kepler field local offset I find of $c=-15\pm2\muas$ (\S\ref{sec:off}). I also note that the quasar parallaxes were not corrected for the $\zf$ model, which will leave spatial systematics due to the ecliptic latitude--dependent zero-point terms modelled with $\zf$, and which could help explain the $\approx2\muas$ larger RMS scatter required to match the quasar and Kepler field covariances. I note \cite{fardal+2021} find that the level of spatial correlation in Gaia DR2 parallaxes increases by a factor of 6 between $G=13$ and $G=20$. Under the assumption that this magnitude dependence is preserved in EDR3 spatial correlations, this would be another reason to expect a larger spatial covariance among quasars than I find here.

The spatial covariance found in this work would be expected to be in better agreement with Gaia
team's estimates of the covariance using the LMC than those using
quasars, since there are not contributions from large-scale
variations. Additionally, the LMC members are more comparable in
magnitude to the sample here. Indeed, the estimate from \cite{l20a} of
a systematic uncertainty of $6.9\muas$ for $0.1\deg \lesssim
\theta_{ij} \lesssim 4.5\deg $ accords better with the present estimate of $4 \pm 1 \muas$ at the smallest scales, keeping in mind the larger systematic expected of the LMC estimate due to not correcting the parallaxes using $\zf$.

I stress that the spatial covariance estimation procedure used here will not be sensitive to any average parallax systematic that persists over the entire Kepler field of view, since I correct the parallaxes according to Equation~\ref{eq:dz}. However, the local offset term that I find, $c=-15 \pm 2\muas$, is a conservative estimate of what sort of average offset the asteroseismic and Gaia parallaxes have, averaged over the whole Kepler field (see \S\ref{sec:off}). This is consistent with the expected RMS of $8.1 \muas$ on scales larger than $10\deg$ using quasars \citep{l20a}. Studies making use of the small-scale covariance as described by Equation~\ref{eq:spatial} and Table~\ref{tab:spatial} would therefore be advised that the large-scale contributions to the covariance may well be larger than the small-scale ones that are available using the Kepler field. 

\section{Concluding remarks}
\label{sec:conc}
I have compared Gaia EDR3 parallaxes with five-parameter solutions to parallaxes derived from knowledge of the temperature, bolometric flux, and asteroseismic radius for more than 2,000 first-ascent red giant branch stars in the Kepler field, and conclude the following:

\begin{enumerate}
    \item The $\zf$ zero-point model \citep{l20b} brings Gaia EDR3 parallaxes into agreement with asteroseismic parallaxes to an average of $\approx 2\muas$ before any adjustments to $\zf$ or asteroseismic parallaxes for the regime $10.8 \leq G < 13$, $1.4 \nueffunit \lesssim \nueff \lesssim 1.5 \nueffunit$ (purple band in Fig.~\ref{fig:vplx}).
    \item I find strong evidence that the $\zf$ zero-point model over-corrects parallaxes for $ 9 \lesssim G \lesssim 11$, such that corrected parallaxes are too large by $15\muas \pm 3\muas$ (purple band in Fig.~\ref{fig:res}f), consistent with the findings for even brighter Cepheids distributed across the sky ($6 \lesssim G \lesssim 10$; \citealt{riess+2020}).
    \item I identify a significant color dependence to the asteroseismic-Gaia parallax residuals of $-300 \pm 25\dunit$ for $G \gtrsim 11$ and $1.4 \nueffunit \lesssim \nueff \lesssim 1.5 \nueffunit$, of which $\sim 100\dunit$ may plausibly be attributed to asteroseismic parallax systematic uncertainties. If this color term were due to unmodelled behavior in $\zf$, there is reason to believe it may have an ecliptic latitude dependence.
    \item After removing the aforementioned color term from the asteroseismic-Gaia parallax residuals, and after adjusting $\zf$ according to an additive $-15\pm 3\muas$ offset for $G< 10.8$, Gaia EDR3 and asteroseismic parallaxes agree to within a constant offset of $c = -15 \pm 2\muas$. The latter may be interpreted as a conservative estimate of the large-scale variation in the parallax zero-point, given its dependence on color trends, and is broadly consistent with the Gaia team's estimates of spatially-correlated parallax variations on scales larger than the Kepler field.
    \item Gaia EDR3 parallax statistical uncertainties in the $9 \lesssim G \lesssim 13$, $1.4 \nueffunit \lesssim \nueff \lesssim 1.5 \nueffunit$ regime probed by the sample are under-estimated by $22\pm6\%$, consistent with estimates from the Gaia team \citep{fabricius+2020} and in good agreement with independent analysis using wide binaries \citep{elbadry}.
    \item The spatial correlations in Gaia EDR3 parallaxes on the scales probed by the Kepler field ($\lesssim 10\deg$) --- described by Equation~\ref{eq:spatial} with parameters in Table~\ref{tab:spatial} --- agree broadly with those inferred by \cite{l20a}, to within an additive constant due to correlations on scales larger than the Kepler field.  These spatial correlations incur a $\approx 4\muas$ systematic parallax uncertainty on angular scales less than $ \approx 0.1 \deg$, falling to below a $\approx 0.1\muas$ level on scales larger than $ \approx 5 \deg$ (Fig.~\ref{fig:spatial}). 
    \end{enumerate}

The zero-point in EDR3 is not expected to be significantly different in the next installment of Data Release 3 scheduled in 2022, since the next complete astrometric solution will only be done for Gaia DR4. At that time, further reduction in the parallax zero-point is expected in part due to the reversal of the Gaia satellite's precession \citep{l20a}. Until such a time, further characterization of the EDR3 $\zf$ model would be fruitful, particularly to definitively constrain any color adjustments needed of it and to validate it as a function of ecliptic latitude.

\startlongtable
\begin{splitdeluxetable*}{cccccccccBccccccccc}
  \tablecaption{Best-fitting Gaia EDR3 parallax zero-point model adjustment parameters \label{tab:res}}
  \tablehead{Model (\#) & $\Delta$BIC$^{a}$ & $c$ [$\muas$] & $c_2$ [$\muas$] & $d$ [$\muas\, \mu m$]  &  $d_1$ [$\muas\, \mu m$] & $d_2$ [$\muas\, \mu m$] & $d_3$ [$\muas\, \mu m^3$] &$d_a$ [$\daunit$] & $e$ [$\muas\,
      \mathrm{mag}^{-1}$] & $e_1$ [$\muas\,
      \mathrm{mag}^{-1}$] & $e_2$ [$\muas\,
      \mathrm{mag}^{-1}$] & $a$ & $k$ & $g$ & $s$ [$\muas$] & $\chi^2/\mathrm{dof}^{c}$  & N}
\startdata 
  (0) & $  0.0$ & $-14.6 \pm  1.8$ & $ -15 \pm    3$ & $-303 \pm  25$ & $ ---$ & $  ---$ & $   ---$ &$  ---$& $---$ & $---$ & $---$  & $0.995 \pm 0.002$ & $ ---$ & $ ---$ & $---$ & $0.760^{*****}$ & $ 2130$ \\ \hline
 $e$ no $a$ (1) & $ -4.7$ & $-13.1 \pm  1.0$ & $ -17 \pm    3$ & $-289 \pm  24$ & $ ---$ & $  ---$ & $   ---$ &$  ---$& $-2.6 \pm 0.9$ & $---$ & $---$  & $  ---$ & $ ---$ & $ ---$ & $---$ & $0.755^{*****}$ & $ 2130$ \\ \hline
 $d_a$ (2) & $  1.1$ & $-12.9 \pm  1.9$ & $ -15 \pm    3$ & $-222 \pm  40$ & $ ---$ & $  ---$ & $   ---$ & $-0.14 \pm 0.06$ & $---$ & $---$ & $---$  & $0.998 \pm 0.002$ & $ ---$ & $ ---$ & $---$ & $0.769^{*****}$ & $ 2130$ \\ \hline
 $e$ (3) & $  2.8$ & $-13.6 \pm  1.8$ & $ -17 \pm    3$ & $-292 \pm  26$ & $ ---$ & $  ---$ & $   ---$ &$  ---$& $-2.4 \pm 1.1$ & $---$ & $---$  & $0.999 \pm 0.003$ & $ ---$ & $ ---$ & $---$ & $0.756^{*****}$ & $ 2130$ \\ \hline
 $d_1d_2$ (4) & $  6.7$ & $-14.6 \pm  1.8$ & $ -19 \pm    5$ & $ ---$ & $-437 \pm 134$ & $ -298 \pm   26$ & $   ---$& $  ---$& $---$ & $---$ & $---$  & $0.995 \pm 0.002$ & $ ---$ & $ ---$ & $---$ & $0.760^{*****}$ & $ 2130$ \\ \hline
 $d_3$ (5) & $  6.9$ & $-14.2 \pm  1.8$ & $ -15 \pm    3$ & $-280 \pm  34$ & $ ---$ & $  ---$ & $  6710 \pm   7518$ &$  ---$&$  ---$& $---$ & $---$ & $0.996 \pm 0.002$ & $ ---$ & $ ---$ & $---$ & $0.766^{*****}$ & $ 2130$ \\ \hline
 $s$ (6) & $  7.7$ & $-14.5 \pm  1.8$ & $ -15 \pm    3$ & $-301 \pm  25$ & $ ---$ & $  ---$ & $   ---$ &$  ---$& $---$ & $---$ & $---$  & $0.996 \pm 0.002$ & $ ---$ & $ ---$ & $  0 \pm   1$ & $0.761^{*****}$ & $ 2130$ \\ \hline
 $e_1e_2$ (7) & $ 10.4$ & $-13.6 \pm  1.9$ & $ -20 \pm   13$ & $-292 \pm  25$ & $ ---$ & $  ---$ & $   ---$ &$  ---$& $---$ & $-4.0 \pm 7.1$ & $-2.4 \pm 1.1$  & $0.999 \pm 0.003$ & $ ---$ & $ ---$ & $---$ & $0.756^{*****}$ & $ 2130$ \\ \hline
 no $c_2$ (8) & $ 14.9$ & $-11.5 \pm  1.7$ & $ ---$ & $-293 \pm  25$ & $ ---$ & $  ---$ & $   ---$ & $  ---$& $---$ & $---$ & $---$  & $1.000 \pm 0.002$ & $ ---$ & $ ---$ & $---$ & $0.766^{*****}$ & $ 2130$ \\ \hline
 $d_a$ no $d$ (9) & $ 22.1$ & $-6.1 \pm  1.5$ & $ -14 \pm    3$ & $ ---$ & $ ---$ & $  ---$ & $   ---$ & $-0.37 \pm 0.03$ & $---$ & $---$ & $---$  & $1.005 \pm 0.002$ & $ ---$ & $ ---$ & $---$ & $0.767^{*****}$ & $ 2130$ \\ \hline
  no $c$ (10) & $ 58.8$ & $---$ & $  -6 \pm    3$ & $-180 \pm  20$ & $ ---$ & $  ---$ & $   ---$ & $---$ & $  ---$& $---$ & $---$  & $1.011 \pm 0.001$ & $ ---$ & $ ---$ & $---$ & $0.797^{*****}$ & $ 2130$ \\ \hline
 no $d$ (11) & $137.8$ & $-1.2 \pm  1.4$ & $ -12 \pm    3$ & $ ---$ & $ ---$ & $  ---$ & $   ---$ &$  ---$& $---$ & $---$ & $---$  & $1.002 \pm 0.002$ & $ ---$ & $ ---$ & $---$ & $0.863^{****}$ & $ 2130$ \\ \hline
 $k$ $g$ (12) & $-105.4$ & $-14.4 \pm  1.6$ & $ -14 \pm    2$ & $-340 \pm  24$ & $ ---$ & $  ---$ & $  ---$&$   ---$ & $---$ & $---$ & $---$  & $0.997 \pm 0.002$ & $1.22 \pm 0.06$ & $0.69 \pm 0.03$ & $---$ & $0.999^{}$ & $ 2130$ \\ \hline
 El-Badry,Rix,Heintz2021 $k$ $g$ (13) & $---^{b}$ & $-15.6 \pm  1.7$ & $ -14 \pm    2$ & $-353 \pm  24$ & $ ---$ & $  ---$ & $   ---$ & $---$ &$  ---$& $---$ & $---$  & $0.996 \pm 0.002$ & $1.02 \pm 0.04$ & $0.70 \pm 0.03$ & $---$ & $0.998^{}$ & $ 2151$ \\ \hline
 Yu+2018 (14) & $---^{b}$ & $-14.5 \pm  1.8$ & $ -17 \pm    3$ & $-182 \pm  27$ & $ ---$ & $  ---$ & $  ---$&$   ---$ & $---$ & $---$ & $---$  & $0.984 \pm 0.002$ & $ ---$ & $ ---$ & $---$ & $0.904^{***}$ & $ 2114$ \\ \hline
 $e$ Yu+2018 (15) & $7.6^{b}$ & $-14.4 \pm  1.8$ & $ -17 \pm    3$ & $-181 \pm  27$ & $ ---$ &$  ---$& $  ---$ & $   ---$ & $0.0 \pm 1.1$ & $---$ & $---$  & $0.984 \pm 0.003$ & $ ---$ & $ ---$ & $---$ & $0.904^{***}$ & $ 2114$ \\ \hline
 Kallinger+2018 (16) & $---^{b}$ & $-15.4 \pm  1.8$ & $ -14 \pm    3$ & $-212 \pm  25$ & $ ---$ & $  ---$ & $   ---$ & $---$ & $  ---$&$---$ & $---$  & $0.978 \pm 0.002$ & $ ---$ & $ ---$ & $---$ & $0.805^{*****}$ & $ 2140$ \\ \hline
 $e$ Kallinger+2018 (17) & $0.6^{b}$ & $-14.3 \pm  1.8$ & $ -17 \pm    3$ & $-201 \pm  25$ & $ ---$ & $  ---$ & $   ---$ $  ---$&& $-2.9 \pm 1.1$ & $---$ & $---$  & $0.983 \pm 0.003$ & $ ---$ & $ ---$ & $---$ & $0.798^{*****}$ & $ 2140$ \\ \hline
\enddata
\tablecomments{Best-fitting parameters for models based on Equation~\ref{eq:second_model}, according to MCMC analysis using the likelihood of Equation~\ref{eq:second_model}. N denotes the number of stars in the sample used for the fit. \\$^{a}$ The difference in the Bayesian Information Criterion between a given model and Model 0, where a smaller $\Delta$BIC indicates a more preferred model. \\$^{b}$ Models 13-17 use data incommensurate with the data used for Models 0-12, so Model 13 has $\Delta$BIC set to 0; Models 14-15 have $\Delta$BIC listed with respect to Model 14; and Models 16-17 with respect to Model 16. \\ $^{c}$ Asterisks indicate the significance of the deviation of $\chi^2/\mathrm{dof}$ from unity, with one asterisk for each $\sigma$, capped at $5\sigma$.}
  \end{splitdeluxetable*}

\begin{splitdeluxetable*}{ccccBccccc}
    \tablecaption{Gaia EDR3 parallax spatially-correlated systematic uncertainties \label{tab:spatial}}                        
  \tablehead{$\rho$ & $\theta_{1/2}$  & $\chi^2/\mathrm{dof}$  & N & $\sqrt{C_{ij}(\theta_{ij} = \theta_{1/2})}$$^a$ & $\sqrt{C_{ij}(\theta_{ij} = 0\deg)}$$^a$ &$\sqrt{C_{ij}(\theta_{ij} = 0.5\deg)}$$^a$ & $
      \sqrt{C_{ij}(\theta_{ij} = 1\deg)}$$^a$ & $\sqrt{C_{ij}(\theta_{ij} = 5\deg)}$$^a$}
\startdata 
  $15^{+9}_{-9}\muas^2$ &  $0.49^{+0.17}_{-0.17}\mathrm{deg}$ &$2.9$  & $37$ & $2.5^{+0.6}_{-0.8}\muas$ & $3.9^{+1.1}_{-1.4}\muas$ & $2.5^{+0.6}_{-0.8}\muas$ & $1.6^{+0.5}_{-0.6}\muas$ & $0.1^{+0.1}_{-0.1}\muas$\\
 \enddata
 \tablecomments{Best-fitting parameters for spatially-correlated uncertainties of Gaia parallaxes, as inferred in the Kepler field and parametrized by Equation~\ref{eq:spatial}. $\rho$ describes the variance at small scales. $\theta_{1/2}$ is the characteristic angular scale for the spatially-correlated uncertainties. \\$^{a}$ The square root of the covariance, $C$, is a measure of the systematic uncertainty in the Gaia parallaxes due to spatial correlations.}
\end{splitdeluxetable*}

\software{asfgrid \citep{asfgrid}, NumPy \citep{numpy}, pandas
  \citep{pandas}, Matplotlib \citep{matplotlib}, IPython
  \citep{ipython}, SciPy \citep{scipy}, isoclassify
  \citep{huber+2017,berger+2020}, corner \citep{corner}}

\acknowledgments
The author would like to thank Adam Riess for encouraging, productive conversations and instructive comments. The author also thanks Ruth Angus and the Center for Computational Astrophysics for providing office space. Thanks go especially to the Gaia team for their revolutionary work. The author is supported by an NSF Astronomy and Astrophysics Postdoctoral Fellowship under award AST-2001869.

This work has made use of data from the European Space Agency (ESA) mission
{\it Gaia} (\url{https://www.cosmos.esa.int/gaia}), processed by the {\it Gaia}
Data Processing and Analysis Consortium (DPAC,
\url{https://www.cosmos.esa.int/web/gaia/dpac/consortium}). Funding for the DPAC
has been provided by national institutions, in particular the institutions
participating in the {\it Gaia} Multilateral Agreement.

Funding for the Sloan Digital Sky 
Survey IV has been provided by the 
Alfred P. Sloan Foundation, the U.S. 
Department of Energy Office of 
Science, and the Participating 
Institutions. 

SDSS-IV acknowledges support and 
resources from the Center for High 
Performance Computing  at the 
University of Utah. The SDSS 
website is www.sdss.org.

SDSS-IV is managed by the 
Astrophysical Research Consortium 
for the Participating Institutions 
of the SDSS Collaboration including 
the Brazilian Participation Group, 
the Carnegie Institution for Science, 
Carnegie Mellon University, Center for 
Astrophysics | Harvard \& 
Smithsonian, the Chilean Participation Group, the French Participation Group, Instituto de Astrof\'isica de 
Canarias, The Johns Hopkins 
University, Kavli Institute for the 
Physics and Mathematics of the 
Universe (IPMU) / University of 
Tokyo, the Korean Participation Group, Lawrence Berkeley National Laboratory, Leibniz Institut f\"ur Astrophysik 
Potsdam (AIP),  Max-Planck-Institut 
f\"ur Astronomie (MPIA Heidelberg), 
Max-Planck-Institut f\"ur 
Astrophysik (MPA Garching), 
Max-Planck-Institut f\"ur 
Extraterrestrische Physik (MPE), 
National Astronomical Observatories of 
China, New Mexico State University, 
New York University, University of 
Notre Dame, Observat\'ario 
Nacional / MCTI, The Ohio State 
University, Pennsylvania State 
University, Shanghai 
Astronomical Observatory, United 
Kingdom Participation Group, 
Universidad Nacional Aut\'onoma 
de M\'exico, University of Arizona, 
University of Colorado Boulder, 
University of Oxford, University of 
Portsmouth, University of Utah, 
University of Virginia, University 
of Washington, University of 
Wisconsin, Vanderbilt University, 
and Yale University.

This publication makes use of data products from the Two Micron All Sky Survey, which is a joint project of the University of Massachusetts and the Infrared Processing and Analysis Center/California Institute of Technology, funded by the National Aeronautics and Space Administration and the National Science Foundation.

\bibliography{bib}

\end{document}